\documentclass[a4paper,11pt,twoside,fleqn]{revtex4-1}
 \usepackage[greek,english]{babel}
 \usepackage[iso-8859-7]{inputenc}
 \usepackage[dvips]{graphicx}
 \usepackage[dvips]{graphics}
 \usepackage{epsfig}
  \usepackage{enumerate}
  \usepackage{mathrsfs}
\usepackage{amsmath}
\usepackage{amsfonts}
\usepackage{amssymb}
\usepackage{amsthm}
\usepackage{graphics}
\usepackage{subfig}
\usepackage{makeidx}
\usepackage{float}
\usepackage{color}
\usepackage{soul}
\usepackage{fancyhdr}
\usepackage{slashed}
\usepackage[toc,page]{appendix}
\usepackage{afterpage}
\usepackage{epstopdf}
\usepackage{array}
\usepackage{booktabs}
\usepackage{multirow}

 \newcommand{\newc}{\newcommand}
 \newc{\ra}{\rightarrow}
 \newc{\lra}{\leftrightarrow}
 \newc{\beq}{\begin{equation}}
 \newc{\eeq}{\end{equation}}
 \newc{\barr}{\begin{eqnarray}}
 \newc{\barra}{\begin{eqnarray*}}
 \newc{\earr}{\end{eqnarray}}
 \newc{\earra}{\end{eqnarray*}}
 \newc{\texa}{\textstyle}
 \newc{\paral}{\parallel}
 \newc{\und}{\underline}
 \newc{\pars}{\partial}
 \newc{\nonu}{\nonumber \\}
 \newc{\nln}{\\ \vspace{2mm}}
 \newc{\HRule}{\rule{\linewidth}{0.5mm}}

\begin{document}
 
 \begin{center}
 {\large \bf
Axisymmetric equilibria with pressure anisotropy \vspace{2mm} \\ and plasma flow
   } \vspace{3mm}
   
   A. Evangelias$^{1,a)}$, G. N. Throumoulopoulos$^{1,a)}$ \vspace{2mm}
 
  { \it
$^1$University of Ioannina, Physics Department,\\
 Section of Astrogeophysics, GR 451 10 Ioannina, Greece }
\vspace{2mm}
 
 { \it
$^{a)}$Electronic Addresses: aevag@cc.uoi.gr,   gthroum@cc.uoi.gr }
\end{center}
 
 \setlength{\baselineskip}{13.0pt}
 \begin{center}
 \HRule
 \end{center}
 
 A generalised Grad-Shafranov equation that governs the equilibrium of an axisymmetric toroidal plasma with anisotropic pressure and incompressible flow of arbitrary direction is derived. This equation includes six free surface functions and recovers known Grad-Shafranov-like equations in the literature as well as the usual static, isotropic one. The form of the generalised equation indicates that pressure anisotropy and flow act additively on equilibrium. In addition, two sets of analytical solutions, an extended Solovev one with  a plasma reaching the separatrix  and an extended Hernegger-Maschke one for a plasma surrounded by a fixed boundary possessing an X-point, are constructed, particularly in relevance to the ITER and NSTX tokamaks. Furthermore, the impacts both of pressure anisotropy and plasma flow on these equilibria are examined. It turns out that depending on the maximum value and the shape of an anisotropy function, the anisotropy can act either paramagnetically or diamagnetically. Also, in most of the cases considered both the anisotropy and the flow have stronger effects on NSTX equilibria than on ITER ones.
 \begin{center}
 \HRule
 \end{center}
  
 \section{Introduction}
It has been established in a number of fusion devices
that sheared flow both zonal and mean (equilibrium)
play a role in the transitions to improved confinement
regimes as the L-H transition and the Internal Transport
Barriers \cite{zonal}, \cite{flow}. These flows can be driven externally
in connection with electromagnetic power and
neutral beam injection for plasma heating and current
drive or can be created spontaneously (zonal flow). An additional effect of external heating, depending on the direction of the injected momentum,  is pressure anisotropy which also may play a role in several magnetic fusion related problems. \par
In many important plasmas as the high temperature ones the collision time is so long that collisions can be ignored. It would appear that for such collisionless plasmas a fluid theory should not  be appropriate. However, for perpendicular motions because of gyromotion the magnetic field plays the role of collisions, thus making a fluid description appropriate. Macroscopic equations for a collisionless plasma with pressure anisotropy have been derived by Chew, Goldberger, and Low \cite{CGL} on the basis of a diagonal pressure tensor consisting of one element parallel to the magnetic field and a couple of identical  perpendicular elements associated with two degrees of freedom. \par
The MHD equilibria of axisymmetric plasmas,  which can be  starting points of 
stability and transport studies, is governed by the well known  Grad-Shafranov 
(GS) equation. The most widely employed analytic solutions of this equation is the Solovev solution \cite{so} and the Hernegger-Maschke solution \cite{hema}, the former corresponding  to toroidal current density non vanishing on the plasma boundary and the  latter to toroidal current density vanishing thereon.  
In the presence of flow the equilibrium satisfies a generalised  Grad-Shafranov 
(GGS) equation together with a Bernoulli equation involving the pressure 
(see for example \cite{moso,ha,T-Th}). 
For compressible flow the GGS equation can be either elliptic or hyperbolic 
depending on the value of a  Mach function  associated with the poloidal velocity. 
Note that the toroidal velocity is inherently incompressible because of 
axisymmetry. In the presence of compressibility the GGS equation is coupled with the Bernoulli 
equation through the density which is not uniform on magnetic surfaces.
For incompressible flow the density becomes a surface quantity and the  GGS 
equation  becomes elliptic and  decouples from the 
Bernoulli equation (see section II). Consequently one has to solve an easier and 
well posed elliptic boundary value problem. In particular for fixed boundaries,  
convergence to the solution is guaranteed under mild requirements of 
monotonicity for the free functions involved in the GGS equation 
\cite{couhi}. For plasmas with anisotropic pressure the equilibrium equations involve a function associated with this anisotropy [Eq. (\ref{sigma function}) below]. To get a closed set of reduced equilibrium  equations an assumption on the functional dependence of this function is required  (cf. \cite{Cotsaftis}-\cite{Kuznetsov} for static equilibria and  \cite{iabo}-\cite{ivma} for stationary ones).\par
%
\par
%
 In this work we derive a new GGS equation  by including both  anisotropic pressure and incompressible flow of arbitrary direction. This equation consists of six arbitrary surface quantities and recovers  known equations as particular cases, as well as the usual GS equation for a static isotropic plasma. Together we obtain a Bernoulli equation for the quantity $\overline{p}$ [Eq. (\ref{bar pressure})], which may be interpreted as an effective isotropic pressure. For the derivation we assume that the function of pressure anisotropy is uniform on magnetic surfaces. In fact, as it will be shown,  for static equilibria as well as for stationary equilibria either with toroidal flow or incompressible flow parallel to the magnetic field, this property of the anisotropy function  follows if
the current density shares the same surfaces with the magnetic field. Then for appropriate choices of the free functions involved we obtain an extended Solovev solution describing configurations with a non-predefined boundary, and  an extended Hernegger-Maschke solution with a fixed boundary possessing an X-point imposed by  Dirichlet boundary conditions.  On the basis of these solutions we construct ITER-like, as well as NSTX and NSTX-Upgrade-like equilibria for arbitrary flow, both diamagnetic and paramagnetic, to examine the impact both of pressure anisotropy and plasma flow on the equilibrium characteristics. The main conclusions are that the pressure anisotropy and the flow act on equilibrium in an additive way, with the anisotropy having a stronger impact than that of the flow. Also the effects of flow and anisotropy are in general more noticeable  in spherical tokamaks than in conventional ones.\par

 The GGS equation for plasmas with pressure anisotropy and flow is derived in section  II. In section III the generalised Solovev and Hernegger-Maschke solutions are obtained and employed to construct  ITER and NSTX pertinent configurations. Then  the impact of anisotropy and flow on equilibrium quantities, as the pressure and current density, 
 are examined in section IV. Section V summarizes the conclusions.  
 \section{The Generalised Grad-Shafranov Equation}
 The ideal MHD equilibrium states of an axially symmetric magnetically confined plasma with incompressible flow and anisotropic pressure are governed by the following set of equations:
 \beq \label{continuity} \vec{\nabla}\cdot (\rho \vec{v})=0\eeq
\beq \label{momentum conservation} \rho (\vec{v}\cdot \vec{\nabla})\vec{v}=\vec{J}\times \vec{B}-\vec{\nabla}\cdot \stackrel{\textstyle\leftrightarrow}{\rm {\mathbb P}}\eeq
\beq \label{ampere} \vec{\nabla}\times \vec{B}=\mu _0\vec{J}\eeq
\beq \label{faraday} \vec{\nabla}\times \vec{E}=0\eeq
\beq \label{gauss} \vec{\nabla}\cdot \vec{B}=0\eeq
\beq \label{ohm} \vec{E}+\vec{v}\times \vec{B}=0\eeq
The diagonal pressure tensor $\stackrel{\textstyle\leftrightarrow}{\rm {\mathbb P}}$, introduced in \cite{CGL}, consists of one element parallel to the magnetic field, $p_{\parallel}$, and two equal perpendicular ones, $p_{\bot}$, and is expressed as
\beq \label{pressure tensor}\stackrel{\textstyle\leftrightarrow}{\rm {\mathbb P}}=p_{\bot}\stackrel{\textstyle\leftrightarrow}{\rm {\mathbb I}}+\frac{\sigma _d}{\mu _0}\vec{B} \vec{B}\eeq
where the dimensionless function
\beq \label{sigma function}\sigma _d =\mu _0\frac{p_{\|}-p_{\bot}}{|\vec{B}|^2} \eeq 
is a measure of the pressure anisotropy.
Particle collisions in equilibrating parallel and perpendicular energies will reduce $\sigma _d $ and therefore  a collision-dominated plasma can be  described accurately by a scalar pressure. However, because of the low collision frequency a high-temperature confined plasma remains for long anisotropic, once anisotropy is induced  by  external heating sources.
\par
At this point we define the quantity
\beq \label{bar pressure}\overline{p}=\frac{p_{\|}+p_{\bot}}{2} \eeq 
which may be interpreted as an effective isotropic pressure, and which should not be confused with the average plasma pressure
\beq \label{average pressure}<p>=\frac{1}{3}Tr(\stackrel{\textstyle\leftrightarrow}{\rm {\mathbb P}})=\frac{p_{\|}+2p_{\bot}}{3}=\overline{p}-\sigma _d \frac{B^2}{6\mu _0}\eeq
On the basis of Eqs. (\ref{sigma function}) and (\ref{bar pressure}), the following instructive relations arise for the two scalar pressures:
\beq \label{perpendicular pressure}p_{\bot}=\overline{p}-\sigma _d \frac{B^2}{2\mu _0}\eeq
and
\beq \label{parallel pressure}p_{\|}=\overline{p}+\sigma _d \frac{B^2}{2\mu _0}\eeq
Owing to axisymmetry, the divergence-free fields, i.e., the magnetic field, the current density, $\vec{J}$,  and the momentum density  of the fluid element, $\rho \vec{v}$,  can be expressed in terms of the stream functions $\psi (R,z)$, $I(R,z)$, $F(R,z)$ and $\Theta (R,z)$ as
\beq \label{magnetic field} \vec{B}=I\vec{\nabla}\phi +\vec{\nabla}\phi \times \vec{\nabla}\psi\eeq
\beq \label{current density} \vec{J}=\frac{1}{\mu _0}(\Delta ^{*}\psi \vec{\nabla}\phi -\vec{\nabla}\phi \times \vec{\nabla}I)\eeq
and
\beq \label{mass flow} \rho \vec{v}=\Theta \vec{\nabla}\phi +\vec{\nabla}\phi \times \vec{\nabla}F\eeq
Here, ($R$, $\phi$, $z$) denote the usual right-handed cylindrical coordinate system;  constant $\psi$ surfaces are the magnetic surfaces; $F$ is related to the poloidal flux of the momentum density field, $\rho \vec{v}$;   the quantity $I=RB_\phi$ is related to the net poloidal current flowing in   the plasma and the toroidal field coils;   $\Theta =\rho R v_{\phi} $;   $\Delta^{*}$ is the elliptic operator defined by $\Delta^{*}\equiv R^2\vec{\nabla}\cdot (\vec{\nabla}/R^2)$; and  $\vec{\nabla}\phi \equiv \hat{e_{\phi}}/R$. \par
 Equations (\ref{continuity})-(\ref{ohm}) can be reduced by means of certain integrals of the system, which are shown to be surface quantities. To identify two of these quantities, the time independent electric field is expressed by $\vec{E}=-\vec{\nabla}\Phi$ and the Ohm's law,  (\ref{ohm}),  is projected along $\vec{\nabla}\phi$ and $\vec{B}$, respectively, yielding
 \beq \label{first integral}\vec{\nabla}\phi \cdot (\vec{\nabla}F \times \vec{\nabla}\psi ) =0\eeq
and
\beq \label{second integral}\vec{B}\cdot \vec{\nabla}\Phi =0\eeq
Equations (\ref{first integral}) and (\ref{second integral}) imply that $F=F(\psi)$ and $\Phi =\Phi (\psi)$. An additional surface quantity is found from the component of Eq. (\ref{ohm}) perpendicular to a magnetic surface:
\beq \label{third integral} \Phi ^{'}=\frac{1}{\rho R^2}(IF^{'}-\Theta )\eeq
where the prime denotes differentiation with respect to $\psi$. On the basis of Eq. (\ref{third integral}) the velocity [Eq. (\ref{mass flow})] can be written in the form
\beq \label{velocity} \vec{v}=\frac{F^{'}}{\rho}\vec{B}-R^2\Phi ^{'}\vec{\nabla}\psi\eeq
Thus, $\vec{v}$ is decomposed into a component parallel to $\vec{B}$ and a non parallel one associated with the electric field in consistence with the Ohm's law (\ref{ohm}). 
Subsequently, by projecting Eq. (\ref{momentum conservation}) along $\vec{\nabla}\phi$ we find a fourth surface quantity of the system:
\beq \label{fourth integral}X(\psi)\equiv (1-\sigma _d -M_p ^2)I+\mu _0R^2F^{'}\Phi^{'}\eeq
Here we have introduced the poloidal Mach function as:
\beq \label{Mach function} M_p ^2\equiv \mu_0\frac{(F^{'})^2}{\rho}=\frac{v_{pol}^2}{B_{pol}^2/\mu_0\rho}=\frac{v_{pol}^2}{v_{Apol}^2}\eeq
where $v_{Apol}=\frac{B_{pol}}{\sqrt{\mu_0\rho}}$ is the Alfv\' en velocity associated with the poloidal magnetic field.
From Eqs. (\ref{third integral}) and (\ref{fourth integral}) it follows that,  neither $I$ is  a surface quantity, unlike the case of static, isotropic equilibria, nor  $\Theta$.
\newline
\newline
With the aid of Eqs. (\ref{first integral})-(\ref{velocity}) and (\ref{fourth integral}), the components of Eq. (\ref{momentum conservation}) along $\vec{B}$ and perpendicular to a magnetic surface are put in the respective forms
\beq \label{Bernoulli independent flow} \vec{B}\cdot \left\lbrace \vec{\nabla}\left[\frac{v^2}{2}+\frac{\Theta \Phi ^{'}}{\rho}\right]+\frac{1}{\rho}\vec{\nabla}\overline{p}\right\rbrace =0\eeq
and
\begin{eqnarray} \label{GS independent flow}
&&\left\lbrace \vec{\nabla} \cdot \left[(1-\sigma _d -M_p ^2)\frac{\vec{\nabla}\psi}{R^2}\right]+
\left[\mu _0 \frac{F^{'}F^{''}}{\rho}-(1-\sigma _d )^{'}\right]\frac{|\vec{\nabla}\psi |^2}{R^2}
-\mu _0 \sigma _d^{'}\frac{B^2}{2\mu _0}\right\rbrace |\vec{\nabla}\psi |^2 \nonumber \\
&+&\left\lbrace  \mu _0 \rho \vec{\nabla}\left(\frac{v^2}{2}\right)-\frac{\mu _0 \rho}{2R^2}\vec{\nabla}\left(\frac{\Theta}{\rho}\right)^2+\frac{(1-\sigma _d )}{2R^2}\vec{\nabla}I^2 +\mu _0 \vec{\nabla}\overline{p}\right\rbrace \cdot \vec{\nabla}\psi =0
\end{eqnarray}
Therefore, irrespective of compressibility the equilibrium is governed by the  equations (\ref{Bernoulli independent flow}) and (\ref{GS independent flow}) coupled through the density, $\rho$, and the pressure anisotropy function, $\sigma_d$. 
Equation (\ref{GS independent flow})  has a singularity when $\sigma_d + M_p ^2 =1$, and so we must assume that 
$\sigma _d + M_p ^2 \neq 1$. \par
In order to reduce the equilibrium equations further, we employ the incompressibility condition
\beq \label{incompressibility condition} \vec{\nabla}\cdot \vec{v}=0\eeq
Then Eq. (\ref{continuity}) implies that the density is a surface quantity, 
\beq \label{density surface function}\rho =\rho (\psi)\eeq
and so is the Mach function
\beq \label{Mach surface function}M_p ^2=M_p ^2(\psi)\eeq
In addition to obtain a closed set of equations  following \cite{Cotsaftis,Clement,Kuznetsov, clst} we assume that $\sigma _d$ is uniform on magnetic surfaces
\beq \label{assumption} \sigma _d =\sigma _d(\psi) \eeq 
For static equilibria this follows from Eq. (\ref{fourth integral}), which becomes $X(\psi)=-I\sigma_d$,  if in the presence of anisotropy the current density  remains on the magnetic surfaces ($I=I(\psi)$).  Since $M_p=M_p(\psi)$, the  same implication for $\sigma_d$ holds for parallel incompressible flow as well as for toroidal flow. Also,  the hypothesis $\sigma _d=\sigma _d(\psi)$, according to \cite{Cotsaftis}, may be the only suitable for satisfying the boundary conditions on a  rigid, perfectly conducting wall. \par
From Eqs. (\ref{third integral}) and (\ref{fourth integral}) it follows that axisymmetric equilibria with purely poloidal flow $(\Theta =0)$ cannot exist because of the following contradiction: from Eq. (\ref{fourth integral}) it follows that  $I=\frac{X(\psi)}{1-\sigma _d (\psi)}$ is a surface function, but also, $I=\frac{\rho (\psi)\Phi ^{'}(\psi)}{F^{'}(\psi)}R^2$ from Eq. (\ref{third integral}), implying that $I$ has an explicit dependence on $R$; so it cannot be a surface function.$  $ On the other hand, there can exist an equilibrium with purely toroidal flow, either ``compressible", in the sense that the density  varies on the magnetic surfaces, or an incompressible one with uniform density $\rho (\psi )$ thereon.
For isotropic plasmas both kinds of these  equilibria were examined in \cite{pothta}.  \par

With the aid of Eq. ({\ref{density surface function}), Eq. (\ref{Bernoulli independent flow}) can be integrated to yield an expression for the effective pressure, i.e.,
\beq \label{Bernoulli} \overline{p}=\overline{p}_s (\psi)-\rho \left[\frac{v^2}{2}-\frac{(1-\sigma _d) R^2 (\Phi ^{'})^2}{1-\sigma _d -M_p ^2}\right]\eeq
 Therefore, in the presence of flow the magnetic surfaces in general do not coincide with the  surfaces on which $\overline{p}$ is uniform. In this respect, the term containing $\overline{p}_s (\psi)$ is the static part of the effective pressure which does not vanish when $\vec{v}=0$. \par
Finally, by inserting  Eq. (\ref{Bernoulli}) into Eq. (\ref{GS independent flow}) after some algebraic manipulations,  the latter reduces to the following elliptic differential equation,

\begin{eqnarray} \label{GGS psi}
&&(1-\sigma _d -M_p ^2)\Delta ^{*}\psi +\frac{1}{2} (1-\sigma _d -M_p ^2)^{'}|\vec{\nabla}\psi |^2
+\frac{1}{2}\left(\frac{X^2}{1-\sigma _d -M_p ^2}\right)^{'} \nonumber \\
&+&\mu _0 R^2 \overline{p}_s ^{'}+
\mu _0 \frac{R^4}{2}\left[\frac{(1-\sigma _d)\rho (\Phi ^{'})^2}{1-\sigma _d -M_p ^2}\right]^{'}=0
\end{eqnarray}
This is the GGS equation that governs the equilibrium for an axisymmetric plasma with pressure anisotropy and incompressible flow. For flow parallel to the magnetic field the $R^4$-term vanishes. For vanishing flow Eq. (\ref{GGS psi}) reduces to the one derived in \cite{Clement}, when the pressure is isotropic it reduces to the one obtained in \cite{T-Th}, and when both anisotropy and flow are absent it reduces to the well known GS equation. Equation (\ref{GGS psi}) contains six arbitrary surface quantities, namely: $X(\psi)$, $\Phi (\psi)$, $\overline{p}_s (\psi)$, $\rho (\psi)$, $M_p ^2 (\psi)$ and $\sigma _d (\psi)$, which can be assigned as functions of $\psi$ to obtain analytically solvable  linear  forms of the equation or from other physical considerations.

\subsection{Isodynamicity}

There is  a special class of static equilibria called isodynamic for which the magnetic field magnitude is a surface quantity ($|\vec{B}|=|\vec{B}(\psi)|$) \cite{Palumbo}. This feature can have beneficial effects on confinement because the grad-$B$ drift vanishes and consequently plasma transport perpendicular  to the magnetic surfaces is reduced. Also, it was proved that the only possible isodynamic equilibrium is axisymmetric \cite{PaBo}. For fusion plasmas the thermal conduction along $\vec{B}$ is fast compared to the heat transport perpendicular to a magnetic surface, so a good assumption is that the parallel temperature is a surface function, $T_{\parallel}=T_{\parallel}(\psi )$. Then, assuming  that the plasma obeys the ideal gas law, it follows that the parallel pressure becomes also a surface function, $p_{\parallel}=p_{\parallel}(\psi )$. \par
With the aid of these assumptions, and on the basis of Eqs.  (\ref{velocity}) and (\ref{Bernoulli independent flow})  it follows that the magnitude of the magnetic field is related with the perpendicular pressure as
\beq \label{iso 1}|\vec{B}|^2=\frac{2G(\psi)}{M_p ^2(\psi)}-\left(p_{\bot}-\rho R^2 (\Phi ^{'})^2\right)\frac{1}{M_p ^2(\psi)}\eeq 
where $G(\psi) \equiv \rho \left[\frac{v^2}{2}+\frac{\Theta \Phi ^{'}}{\rho}\right]+\frac{p_{\bot}}{2}$.
We note that $|\vec{B}|^2$ becomes a surface function when the perpendicular pressure satisfies the relation $p_{\bot}=\rho R^2 (\Phi ^{'})^2$. This implies that 
\beq \label{iso 2} \sigma _d=\sigma _d(\psi ,R)=\mu _0 \frac{p_{\parallel} (\psi)}{|\vec{B}|^2(\psi)}-R^2\mu _0\frac{\rho(\psi)(\Phi ^{'})^2(\psi)}{|\vec{B}|^2(\psi)}\eeq
which is in contradiction with the hypothesis that the function $\sigma _d$ is a surface quantity. 
Consequently, the only possibility for isodynamic magnetic surfaces to exist is that for field aligned flow, $\Phi ^{'}=0$, 
because then Eq. (\ref{iso 1}) reduces to 
\beq \label{iso 3}|\vec{B}|^2=\frac{2G(\psi)}{M_p ^2(\psi)}-\frac{p_{\bot}}{M_p ^2(\psi)}\eeq
Eqs. (\ref{sigma function}) and (\ref{iso 3})
 imply that both $|\vec{B}|^2=|\vec{B}|^2(\psi)$ and 
$p_{\bot}=p_{\bot}(\psi)$. \par
Thus, the conclusions for the isotropic case \cite{T-Th} are generalised for 
anisotropic pressure, i.e. all three $B$, $p_{\parallel}$ and $p_{\bot}$ become surface quantities. We note here that the more physically pertinent case that $B$ and $p_{\bot}$ remain arbitrary functions would require either compressibility or eliminating the assumption $\sigma _d=\sigma _d(\psi)$. However, in this case tractability is lost and the problem requires numerical treatment.

\subsection{Generalised Transformation}
Using the transformation  
\beq \label{transformation}u(\psi)=\int_{0}^{\psi} \sqrt{1-\sigma _d (g)-M_p^2 (g)}dg,\qquad \sigma _d +M_p ^2<1\eeq
 Eq. (\ref{GGS psi}) reduces to 
\beq \label{GGS u} \Delta ^{*}u+\frac{1}{2}\frac{d}{du}\left(\frac{X^2}{1-\sigma _d -M_p ^2}\right)+\mu _0 R^2\frac{d\overline{p}_s}{du}+\mu _0 \frac{R^4}{2}\frac{d}{du}\left[(1-\sigma _d )\rho \left(\frac{d\Phi}{du}\right)^2\right]=0\eeq
Transformation (\ref{transformation}) does not affect the magnetic surfaces, it just relabels them by the flux function $u$, and is a generalisation of that introduced in \cite{Simintzis} for isotropic equilibria with incompressible flow ($\sigma_d=0$) and that introduced in \cite{Clement} for static anisotropic equilibria ($M_p^2=0$). 
Note that no quadratic term as $|\vec{\nabla}u|^2$ appears anymore in (\ref{GGS u}). Once a solution of this equation is found, the equilibrium can be completely constructed  with calculations in the $u$-space by using (\ref{transformation}) and the inverse transformation 
\beq \label{inv_transformation}\psi(u)=\int_{0}^{u} (1-\sigma _d (g)-M_p^2 (g))^{-1/2}dg \eeq
\par
Before continuing to the construction of analytical solutions, we find it convenient to make a  normalization by  introducing the dimensionless quantities: $ \xi =R/R_i ,\,\zeta =z/R_i
,\, \widetilde{\overline{p}}=\overline{p}/(B_i ^2 /\mu _0),\,\widetilde{\rho}=\rho/\rho _i
,\, \widetilde{u}=u/B_i R_i ^2,\, \widetilde{I}=I/B_i R_i
,\, \widetilde{\vec{E}}=\vec{E}/v_{A_i}B_i,\, \widetilde{\vec{B}}=\vec{B}/B_i
,\, \widetilde{\vec{J}}=\vec{J}/(B_i /\mu _0 R_i),\, \widetilde{\vec{v}}=\vec{v}/v_{A_i}
$. The index $i$ can be  either $a$ or 0, where $a$ denotes the magnetic axis, and 0 the geometric center of a 
configuration. Thus, the normalization constants  are defined as follows: $R_i$ is the radial coordinate of the configuration's magnetic axis/geometric center, and $B_i$, $\rho _i$, $v_{A_i}=\frac{B_i}{\sqrt{\mu _0 \rho _i}}$ are the magnitude of the magnetic field, the  plasma density, and the Afv\' en velocity thereon. Consequently, with the use of the generalised transformation (\ref{transformation}),  Eqs. (\ref{magnetic field})-(\ref{mass flow}), (\ref{fourth integral}), (\ref{Bernoulli}), and (\ref{GGS u}), are put in the following normalized forms in $u$-space:

\beq \label{normalized magnetic field}\widetilde{\vec{B}}=\widetilde{I}\widetilde{\vec{\nabla}}\phi +(1-\sigma _d -M_p ^2)^{-1/2}\widetilde{\vec{\nabla}}\phi \times \widetilde{\vec{\nabla}}\widetilde{u}\eeq

\beq \label{normalized velocity}\widetilde{\vec{v}}=\frac{M_p}{\sqrt{\widetilde{\rho}}}\widetilde{\vec{B}}-\xi ^2 (1-\sigma _d -M_p ^2)^{1/2}\left(\frac{d\widetilde{\Phi}}{d\widetilde{u}}\right)\widetilde{\vec{\nabla}}\phi \eeq

\begin{eqnarray} \label{normalized current density}
\widetilde{\vec{J}}&=&\left[(1-\sigma _d -M_p ^2)^{-1/2}\widetilde{\Delta ^{*}}\widetilde{u}-\frac{1}{2}(1-\sigma _d -M_p ^2)^{-3/2}\frac{d}{d\widetilde{u}}(1-\sigma _d -M_p ^2)|\widetilde{\vec{\nabla}}\widetilde{u}|^2\right]\widetilde{\vec{\nabla}}\phi \nonumber \\
&-&\widetilde{\vec{\nabla}}\phi \times \widetilde{\vec{\nabla}}\widetilde{I}
\end{eqnarray}

\beq \label{normalized X}\widetilde{X}=(1-\sigma _d -M_p ^2)\left[\widetilde{I}+\xi ^2\left(\frac{d\widetilde{F}}{d\widetilde{u}}\right)\left(\frac{d\widetilde{\Phi}}{d\widetilde{u}}\right)\right]\eeq

\beq \label{normalized Bernoulli}\widetilde{\overline{p}}=\widetilde{\overline{p}_s} (\widetilde{u})-\widetilde{\rho} \left[\frac{\widetilde{v}^2}{2}-(1-\sigma _d) \xi ^2 \left(\frac{d\widetilde{\Phi}}{d\widetilde{u}}\right)^2\right]\eeq

and

\beq \label{normalized GGS} \widetilde{\Delta ^{*}}\widetilde{u}+\frac{1}{2}\frac{d}{d\widetilde{u}}\left(\frac{\widetilde{X}^2}{1-\sigma _d -M_p ^2}\right)+\xi ^2\frac{d\widetilde{\overline{p}_s}}{d\widetilde{u}}+\frac{\xi ^4}{2}\frac{d}{d\widetilde{u}}\left[(1-\sigma _d )\widetilde{\rho} \left(\frac{d\widetilde{\Phi}}{d\widetilde{u}}\right)^2\right]=0 \eeq
where $\widetilde{\Delta ^{*}}=\frac{\partial ^2}{\partial \xi ^2}+\frac{\partial ^2}{\partial \zeta ^2}-\frac{1}{\xi}\frac{\partial}{\partial \xi}$. In  section III for appropriate 
choices of the surface functions, Eq. (\ref{normalized GGS}) will be linearised and solved analytically.

\subsection{Plasma beta and safety factor}

The safety factor, measuring  the rate of change of toroidal flux with respect to poloidal flux through an infinitesimal annulus between two neighboring flux surfaces, is given by the following expression 
\beq q\equiv \frac{d\psi _{tor}}{d\psi _{pol}}=\frac{1}{2\pi}\oint \frac{Idl}{R|\vec{\nabla}\psi |}\eeq
Expressing the length element $dl$ in Shafranov coordinates $(r,\theta )$ \cite{Fr} the above formula becomes
\beq \label{safety factor psi}q=\frac{1}{2\pi}\int_{0}^{2\pi}\frac{I\sqrt{r^2+\left(\frac{\psi _{\theta}}{\psi _{r}}\right)^2}}{R|\vec{\nabla}\psi |}d\theta \eeq
where, $\psi _{\theta}=\frac{\partial \psi}{\partial \theta}$, and $\psi _{r}=\frac{\partial \psi}{\partial r}$.
On the basis of the generalised transformation (\ref{transformation}) and the adopted normalization, Eq. (\ref{safety factor psi}) is put in the following form
\beq q=\frac{1}{2\pi}\int_{0}^{2\pi}\frac{\widetilde{I}(\widetilde{u},\xi)\sqrt{\widetilde{r}^2+\left(\frac{\widetilde{u} _{\theta}}{\widetilde{u} _{r}}\right)^2}}{(1-\sigma _d -M_p ^2)^{-1/2}\xi|\widetilde{\vec{\nabla}}\widetilde{u} |}d\theta \eeq
from which we can calculate numerically the safety factor profile.
\par
For the local value of the safety factor on the magnetic axis  there exists the simpler  analytic  expression
\beq \label{safety factor on axis}q_a =(1-\sigma _d-M_p ^2)^{1/2}\frac{\widetilde{I}}{\xi}\left\lbrace \frac{\partial ^2\widetilde{u}}{\partial \xi ^2}\frac{\partial ^2\widetilde{u}}{\partial \zeta ^2}\right\rbrace _{\xi =\xi _a,\zeta =\zeta _a} ^{-1/2}\eeq
 obtained by expansions of the flux function close to the magnetic axis.
From equations (\ref{normalized X}) and (\ref{safety factor on axis}), one observes that when the flow is parallel to the magnetic field, $\frac{d\Phi }{du}=0 $, then the value of $q_a$ has no dependence on the anisotropy. Indeed $q_a$ becomes independent on $\sigma _{d_a}$, where $\sigma _{d_a}$ is the local value of the anisotropy function on the magnetic axis, $\sigma _{d_a}=\sigma _d|_{R=R_a,z=z_a}$.\par
In the case of anisotropic pressure we represent the plasma pressure by the effective pressure, so that the plasma beta can be defined as
\beq \label{plasma beta}\beta \equiv \frac{\overline{p}}{B^2/2\mu _0}\eeq
It also useful to define separate toroidal and poloidal quantities $\beta$ measuring  confinement efficiency of each component of the magnetic field. The toroidal beta to be used here is 
\beq \beta _t =\frac{\overline{p}}{B_0 ^2/2 \mu _0}\eeq
 Recent experiments on the National Spherical Torus Experiment (NSTX) have made significant progress in reaching high toroidal beta $\beta _t \leq 35\% $ \cite{Menard}, while on ITER the beta parameter is expected to take low values, $\beta _t \sim 2\% $ \cite{ITERbeta}.
\section{ANALYTIC EQUILIBRIUM SOLUTIONS}

\subsection{Solovev-like solution}
According to the Solovev ansatz, 
the free function terms  in the GGS equation are chosen to be linear in $\widetilde{u}$ as
\begin{align} \label{Solovev ansatz}
\widetilde{\overline{p}}_s=\widetilde{\overline{p}}_{s_a}\left(1-\frac{\widetilde{u}}{\widetilde{u}_b}\right),\qquad \widetilde{u}\geq 0 \nonumber \\
\frac{\widetilde{X}^2}{1-\sigma _d-M_p ^2}=\frac{2\epsilon \widetilde{\overline{p}}_{s_a}}{(1+\delta ^2)}\frac{\widetilde{u}}{\widetilde{u}_b}+1 \nonumber \\
\widetilde{\rho} (1-\sigma _d)\left(\frac{d\widetilde{\Phi}}{d\widetilde{u}}\right)^2=\frac{2\lambda \widetilde{\overline{p}}_{s_a}}{(1+\delta ^2)}\left(1-\frac{\widetilde{u}}{\widetilde{u}_b}\right)
\end{align}
Here,  $a$ denotes the magnetic axis and $b$ the plasma boundary; $\delta$ determines the elongation of the magnetic surfaces near the magnetic axis; for $\epsilon >0\ (<0)$ the plasma is diamagnetic (paramagnetic); and $\lambda$ is a non-negative parameter related with the non-parallel component of the flow.  In addition, we impose that the solution $\widetilde{u}$ vanishes on the magnetic axis, $\widetilde{u}_a=0$. \par
With this linearising ansatz the GGS equation (\ref{normalized GGS}) reduces to
\beq \label{linearised GGS Solovev}\widetilde{\Delta ^*}\widetilde{u}+\frac{\widetilde{\overline{p}}_{s_a}}{\widetilde{u}_b}\left[\frac{\epsilon}{(1+\delta ^2)}-\xi ^2-\xi ^4 \frac{\lambda}{(1+\delta ^2)}\right]=0\eeq
which admits the following generalised Solovev solution valid for arbitrary $\widetilde{\rho}$, $\sigma_d$ and $M_{p}^2$:
\beq \label{generalised Solovev solution}\widetilde{u}(\xi ,\zeta)=\frac{\widetilde{\overline{p}}_{s_a}}{2(1+\delta ^2)\widetilde{u}_b}\left[\zeta ^2(\xi ^2-\epsilon)+\frac{\delta ^2+\lambda}{4}(\xi ^2 -1)^2+\frac{\lambda}{12}(\xi ^2 -1)^3\right]\eeq
This solution does not include enough free parameters  to impose desirable  boundary conditions, but has the property that a separatrix is spontaneously formed. Thus,  we can predefine the position of the magnetic axis, $(\xi _a =1,\zeta_a=0)$, chosen as normalization point and the plasma extends from the magnetic axis up to a closed magnetic surface which we will choose to coincide with the separatrix.
\par
For an up-down symmetric (about the midplane $\zeta=0$) magnetic surface, its shape can be  characterized by four parameters, namely, the $\xi$ coordinates of the innermost and outermost points on the midplane, $\xi_{in}$ and $\xi_{out}$, and the $(\xi,\zeta)$ coordinates of the highest (upper) point of the plasma boundary, $(\xi_{up},\zeta_{up})$ (see Fig. 14). In terms of these four parameters we can define the normalized major radius
\beq \label{major radius} \xi _0=\frac{\xi _{in}+\xi _{out}}{2}\eeq
which is the radial coordinate of the geometric center, the minor radius
\beq \label{minor radius}\widetilde{\alpha} =\frac{\xi _{out}-\xi _{in}}{2}\eeq
the triangularity of a magnetic surface
\beq \label{triangularity}t=\frac{\xi _0-\xi _{up}}{\widetilde{\alpha }}\eeq
defined as the horizontal distance between the geometric center and the highest point of the magnetic surface normalized  with respect to minor radius, and the elongation of a magnetic surface
\beq \label{elongation}\kappa =\frac{\zeta _{up}}{\widetilde{\alpha }}\eeq
Usually, we specify the values of $R_0$, $\alpha$, $t$, and $\kappa$, instead of ($\xi _{in}$, $\xi_ {out}$, $\xi _{up}$, $\zeta _{up}$) to characterize the shape of the outermost magnetic surface. On the basis of solution (\ref{generalised Solovev solution}) the latter quantities  can be expressed in terms of $\epsilon$, $\delta$, $\lambda$. Subsequently, in order to make an estimate of realistic values for the free parameters $\epsilon$, $\delta$  and the radial coordinate of the magnetic axis $R_a$ in connection with the tokamaks under consideration,
we employ the relations (\ref{major radius})-(\ref{elongation}) to find  ($\epsilon$, $\delta$ and $R_a$)  in terms of  the known parameters  ($R_0$, $\alpha$, $t$ and $\kappa$).  (For ITER: $R_0=6.2\,m$, $\alpha=2.0\,m$, $\kappa=1.7$, $t=0.33$ / for NSTX: $R_0=0.85\,m$, $\alpha=0.67\,m$, $\kappa=2.2$, $t=0.5$). In the static limit ($\lambda=0$)  this estimation procedure can be performed    analytically and when the plasma is diamagnetic we find 
\begin{align} \label{diamagnetic free parameters}
\epsilon =\frac{(R_0-\alpha)^2}{R_0^2+\alpha ^2} \nonumber \\
\delta =\kappa \sqrt{\frac{\alpha}{R_0}} \nonumber \\
R_a=\sqrt{R_0^2+\alpha ^2}
\end{align}
Also, for a diamagnetic equilibrium it holds  $t=1$ (cf. Fig. 1).
The respective relations  for a paramagnetic equilibrium \cite{Arapoglou} for which $\xi_{in}=0$ are
\begin{align} \label{paramagnetic free parameters}
\epsilon =\frac{(t-1)^4}{4(t^2-2t-1} \nonumber \\
\delta =\frac{\kappa \sqrt{2}}{\sqrt{-t^2+2t+1}} \nonumber \\
R_a=\sqrt{2}R_0
\end{align}
Relations (\ref{diamagnetic free parameters})-(\ref{paramagnetic free parameters}) will also be  employed to assign  values of the free parameters $\epsilon$, $\delta$ and $R_a$  for  non parallel flows ($\lambda \neq 0)$ because in this case the above estimation procedure becomes complicated. 
%
 Afterwards, since the vacuum magnetic field at the geometric center of a configuration is known (ITER: $B_0=5.3\,T$ / NSTX: $B_0=0.43\,T$), we can also estimate its value on the magnetic axis by using the relation $B_a=B_0\frac{R_0}{R_a}$, and therefore the value of $\widetilde{\overline{p}}_{s_a}$ from the relation
 $ \widetilde{\overline{p}}_{s_a}=\frac{\overline{p}_{s_a}}{B_a^2/\mu _0}$ ,
once the maximum pressure for each device is known (ITER: $\sim 10^6 \,Pa$ / NSTX: $\sim 10^4 \,Pa$).
\newline
\newline
Thus, we can fully determine the solution $\widetilde{u}$ from Eq. (\ref{generalised Solovev solution}), 
as well as the position of the characteristic points of the boundary and obtain the ITER-like and NSTX-like, diamagnetic and paramagnetic configurations, whose poloidal cross-section with  a set of magnetic surfaces are shown in Figs. (1)-(3). 
\begin{figure} \label{ITERDIAM}
  \centering
    \includegraphics[width=3.5in]{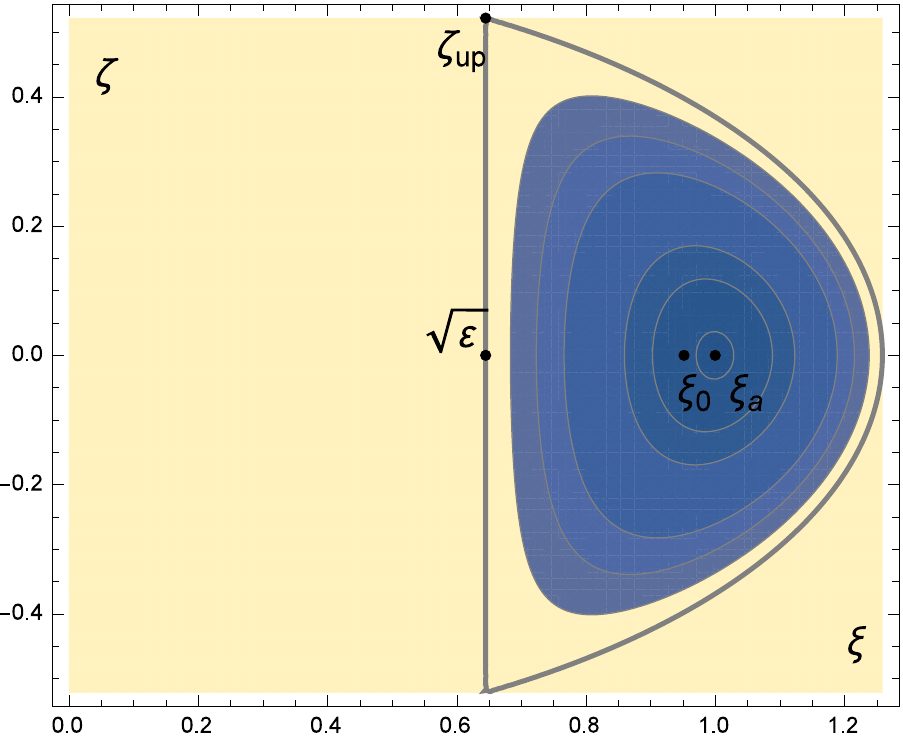}
     \caption{\textit{The static or parallel flow diamagnetic configuration ($\lambda=0$) with ITER-like characteristics corresponding to $\epsilon=0.42$, $\delta=0.97$, $\widetilde{\overline{p}}_{s_a}=0.049$, $\xi_0=0.95$, $\xi_{in}=0.64$, $\xi_{out}=1.26$ and $\widetilde{\alpha}=0.32$.}}
\end{figure}

\begin{figure}\label{NSTXDIAM}
  \centering
    \includegraphics[width=2.0in]{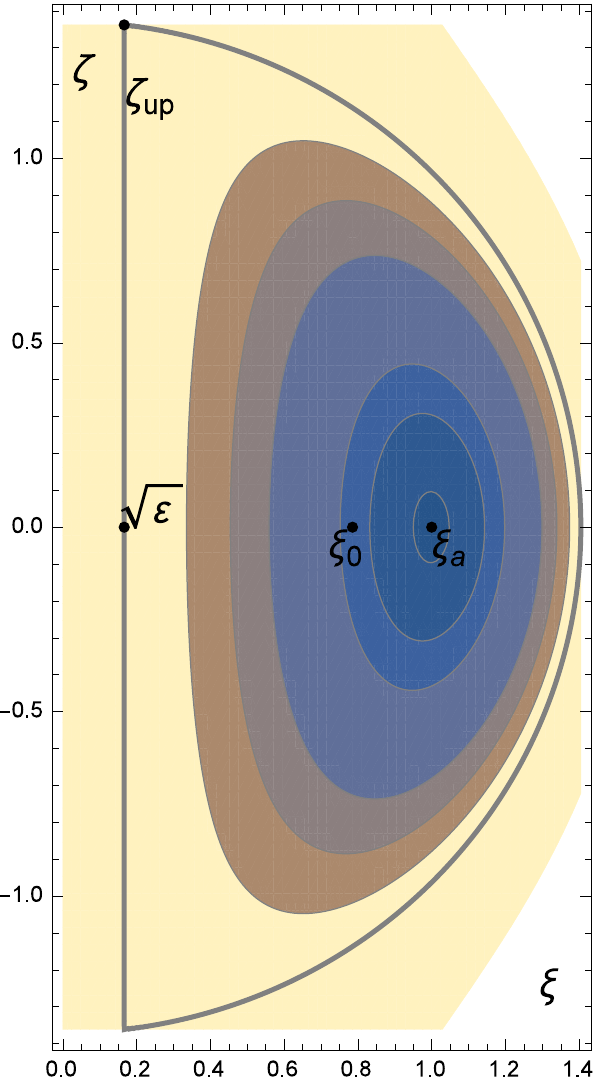}
     \caption{\textit{The static or parallel flow diamagnetic configuration ($\lambda=0$) with NSTX-like characteristics corresponding to $\epsilon=0.03$, $\delta=1.95$, $\widetilde{\overline{p}}_{s_a}=0.11$, $\xi_0=0.78$, $\xi_{in}=0.17$, $\xi_{out}=1.40$ and $\widetilde{\alpha}=0.62$.} }
  \end{figure}
     
\begin{figure} \label{NSTXPAR}
    \includegraphics[width=1.8in]{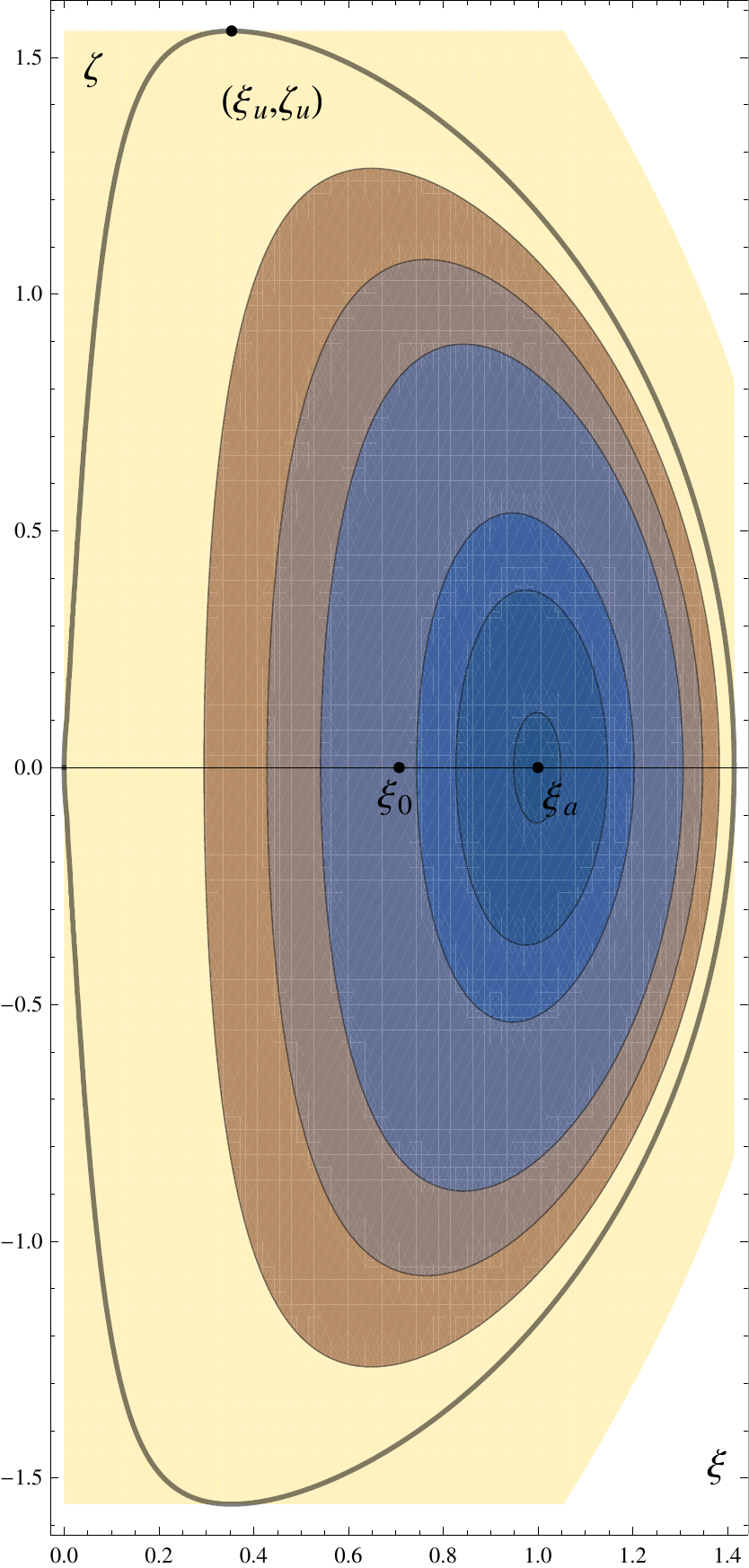}
     \caption{\textit{The static or parallel flow paramagnetic configuration ($\lambda=0$) with NSTX-like characteristics corresponding to $\epsilon=-0.0089$, $\delta=2.35$, $\widetilde{\overline{p}}_{s_a}=0.136$, $\xi_0=1/\sqrt{2}$, $\xi_{in}=0$, $\xi_{out}=\sqrt{2}$ and $\alpha=0.85\,m$.}}    
\end{figure}

We note that by expansions around the magnetic axis it turns out that the magnetic surfaces in the vicinity of the magnetic axis have elliptical cross-sections (see also \cite{Lao}-\cite{Bizarro}). In the diamagnetic configurations presented in Figs. (1) and (2) the inner part of the separatrix is defined by the vertical line $\xi=\sqrt{\epsilon}$, and, for the NSTX  it is located very close to the tokamak axis of symmetry in accordance with the small hole of spherical tokamaks. It may be noted that such D-shaped configurations
are advantageous for improving stability with respect to the interchange modes because of the smaller curvature on the high field side. On the other hand, in a paramagnetic configuration the plasma reaches through a corner the axis of symmetry implying values for the minor radius different from the actual ones, and thus, such a configuration is not typical for conventional tokamaks.  However,   a configuration with a similar   corner   was observed recently in the QUEST spherical tokamak as a self organized state \cite{mizu} (Fig. 5 therein).

\subsection{Hernegger-Maschke-like solution}
 Since the charged particles  move  parallel to the magnetic field free of magnetic force,  parallel flows is a plausible approximation. In particular for tokamaks this is compatible with the fact that the toroidal magnetic field is an order of magnitude larger than the poloidal one and the same scaling is valid  for the toroidal  and poloidal components of the fluid velocity. Also, for parallel flows the problem remains analytically tractable and leads to a generalised Hernegger-Maschke solution to be constructed below. 

In the absence of the electric field term ($\xi^4$ -term) the GGS equation (\ref{normalized GGS}) becomes
\beq \label{static normalized GGS} \widetilde{\Delta ^{*}}\widetilde{u}+\frac{1}{2}\frac{d}{d\widetilde{u}}\left(\frac{\widetilde{X}^2}{1-\sigma _d -M_p ^2}\right)+\xi ^2\frac{d\widetilde{\overline{p}_s}}{d\widetilde{u}}=0\eeq
where all quantities have now been normalized with respect to the geometric center. Choosing the free function terms  of Eq. (\ref{static normalized GGS}) to be quadratic in $\widetilde{u}$ as 
\begin{align} \label{H-M ansatz}
\widetilde{\overline{p}}_s(\widetilde{u})=\widetilde{p}_2\widetilde{u}^2 \nonumber \\
\frac{\widetilde{X}^2(\widetilde{u})}{1-\sigma _d(\widetilde{u})-M_p^2(\widetilde{u})}=1+\widetilde{X}_1\widetilde{u}^2
\end{align}
it reduces to the following linear  differential equation
\beq \label{linearised static normalized GGS}\frac{\partial ^2\widetilde{u}}{\partial \xi ^2}+\frac{\partial ^2\widetilde{u}}{\partial \zeta ^2}-\frac{1}{\xi}\frac{\partial \widetilde{u}}{\partial \xi}+\widetilde{X}_1\widetilde{u}+2\widetilde{p}_2\xi ^2\widetilde{u}=0\eeq
The values of the parameters $\widetilde{p}_2$ and $\widetilde{X}_1$, will be chosen in connection with realistic shaping and values of the equilibrium figures of merit, i.e. the local toroidal beta and the safety factor on the magnetic axis. The solution to Eq. (\ref{linearised static normalized GGS}) is found by separation of variables,
\beq \widetilde{u}(\xi ,\zeta )=G(\xi)T(\zeta)\eeq
on the basis of which, it further reduces to the following form
\beq \frac{1}{T(\zeta)}\frac{d^2T(\zeta)}{d\zeta ^2}=-\frac{1}{G(\xi)}\frac{d^2G(\xi)}{d\xi ^2}+\frac{1}{\xi G(\xi)}\frac{dG(\xi)}{d\xi}-\widetilde{X}_1-2\widetilde{p}_2\xi ^2=-\eta ^2\eeq  
where $\eta$ is the separation constant. \par
 Therefore, the problem reduces to a couple of ODEs. The one for the function $T$ is
\beq  \frac{d^2T(\zeta)}{d\zeta ^2}+\eta ^2T(\zeta)=0\eeq
having the general solution
\beq T(\zeta)=a_1cos(\eta \zeta) +a_2sin(\eta \zeta)\eeq
with the coefficients $a_1$ and $a_2$ to be determined later. The second equation satisfied by the function $G$ is 
\beq \label{eq for G 1}\frac{d^2G(\xi)}{d\xi ^2}-\frac{1}{\xi }\frac{dG(\xi)}{d\xi}+(\widetilde{X}_1-\eta ^2)G(\xi)+2\widetilde{p}_2\xi ^2 G(\xi)=0\eeq
Introducing the parameters $\gamma =\widetilde{X}_1$, $\delta =2 \widetilde{p}_2$, and $\varrho =i\sqrt{\delta}\xi ^2$, so that $\frac{\partial}{\partial \xi}=2i\sqrt{\delta}\xi\frac{\partial}{\partial \varrho}$ and $\frac{\partial^2}{\partial \xi^2}=2i\sqrt{\delta}\frac{\partial}{\partial \varrho}+4i\sqrt{\delta}\varrho\frac{\partial^2}{\partial \varrho^2}$, Eq. (\ref{eq for G 1}) becomes
\beq \label{eq for G 2}\frac{d^2G(\varrho)}{d\varrho ^2}+\left[i\frac{\eta ^2-\gamma}{4\sqrt{\delta}}\frac{1}{\varrho}-\frac{1}{4}\right]G(\varrho)=0\eeq
Furthermore, if we set
$ \nu \equiv i\frac{\eta ^2-\gamma}{4\sqrt{\delta}}$
then Eq. (\ref{eq for G 2}) is put in the form
\beq \frac{d^2G(\varrho)}{d\varrho ^2}+\left[\frac{\nu}{\varrho}-\frac{1}{4}\right]G(\varrho)=0\eeq
which is a special case of the Whittaker's equation for $\mu=\frac{1}{2}$, and thus, it admits the general solution
\beq G(\varrho)=b_1M_{\nu ,\frac{1}{2}}(\varrho)+b_2W_{\nu ,\frac{1}{2}}(\varrho)\eeq
Here,  $M_{\nu,\mu}$ and $W_{\nu,\mu}$ are the Whittaker functions, which are independent solutions of the homonemous differential equation. Consequently, a typical solution of the original equation (\ref{linearised static normalized GGS}) is written in the form
\beq \widetilde{u}(\varrho ,\zeta)=\left[b_1M_{\nu ,\frac{1}{2}}(\varrho)+b_2W_{\nu ,\frac{1}{2}}(\varrho)\right]\left[a_1cos(\eta \zeta) +a_2sin(\eta \zeta)\right]\eeq
For further treatment it is convenient to restrict the separation constant $\eta$ to positive integer values $j$. Therefore, by superposition the solution can be expressed as

{\small
\beq \label{HM u} \widetilde{u}(\varrho ,\zeta)=\sum _{j=1}^\infty \left[a_jM_{\nu _j,\frac{1}{2}}(\varrho)cos(j\zeta)+b_jM_{\nu _j,\frac{1}{2}}(\varrho)sin(j\zeta)+c_jW_{\nu _j,\frac{1}{2}}(\varrho)cos(j\zeta)+d_jW_{\nu _j,\frac{1}{2}}(\varrho)sin(j\zeta)\right]\normalsize \eeq
\normalsize}

Following the analysis given in Appendix A we fully specify the solution (\ref{HM u}) and construct the diverted equilibrium with ITER-like characteristics shown in Fig. (4). Note that the magnetic axis is located outside of the midplane $\zeta=0$ at $(\xi_a=1.05815,\zeta_a=0.0159088)$.

\begin{figure}
  \centering
    \includegraphics[width=3.2in]{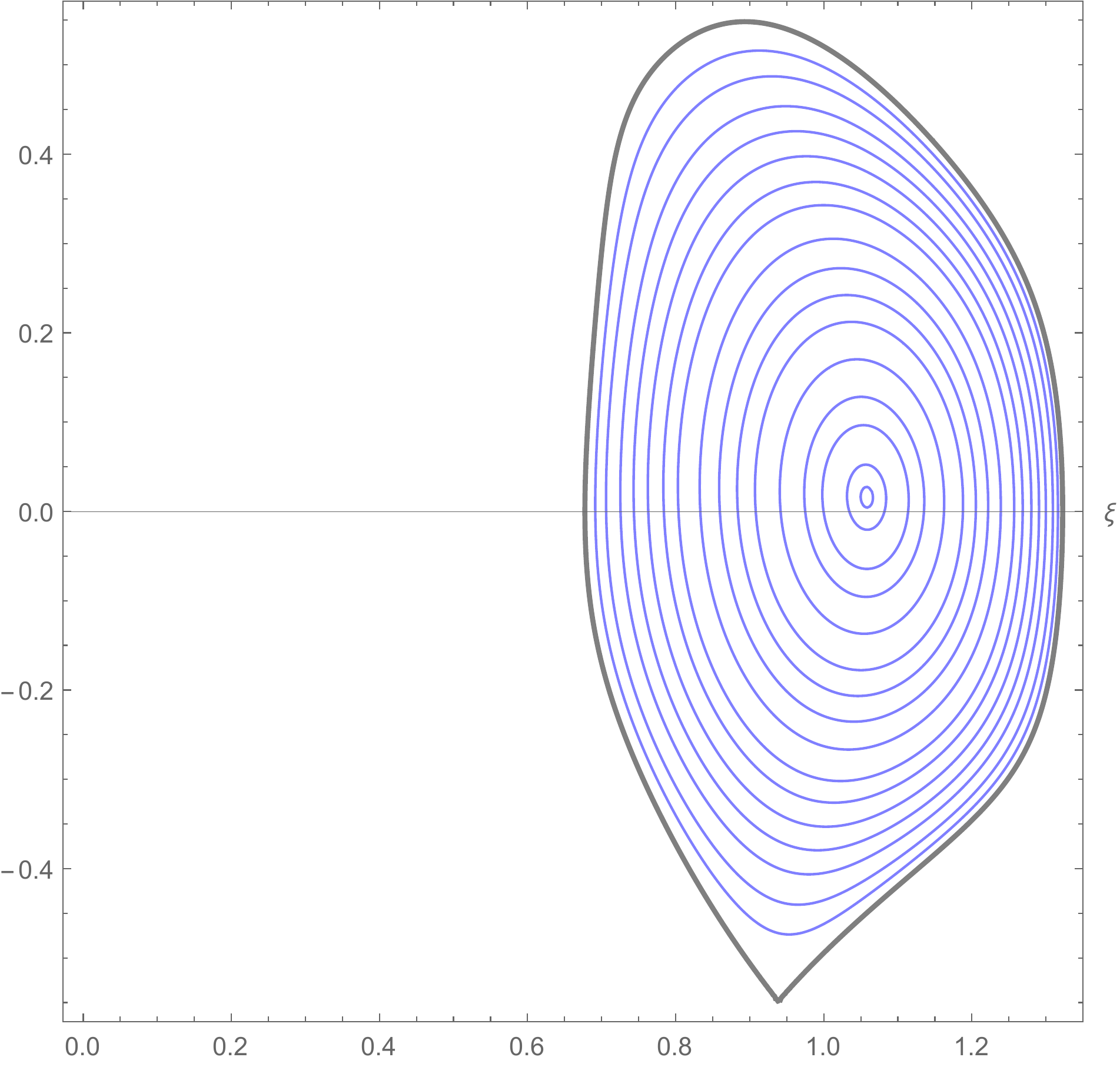}
     \caption{\textit{The poloidal cross-section for an ITER-like diamagnetic equilibria with a lower X-point on the basis of Hernegger-Maschke solution.} }
     \end{figure}
In a similar way we constructed NSTX-Upgrade-like equilibria ($R_0=0.93\,m$, $\alpha=0.57\,m$, $B_0=1.0\,T$, $\kappa=2.5$, and $t=0.3$) on the basis of the extended Hernegger-Maschke solution,  shown in Fig. (5). The magnetic axis of the respective configuration is located at the position $(\xi_a=1.19012,\zeta_a=0.0511745)$, while the flux function on axis takes the value $\widetilde{u}_a^*=0.948259$.
\begin{figure}
  \centering
    \includegraphics[width=2.8in]{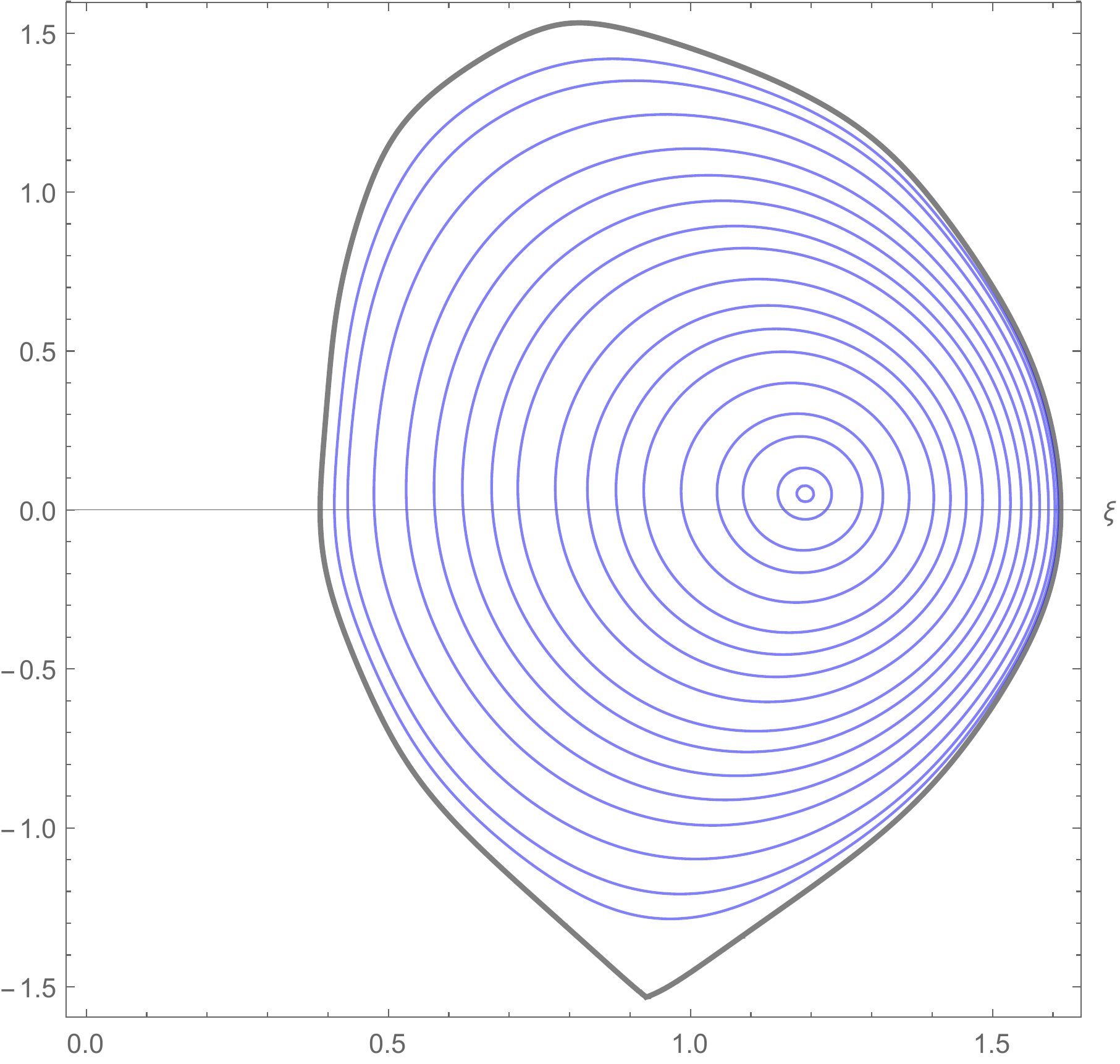}
     \caption{\textit{An NSTX-U-like diamagnetic equilibrium configuration for values of the free parameters $\widetilde{p}_2=4.825$, $\widetilde{X}_1=-0.65$.}}   
     \end{figure}
\section{EFFECTS OF ANISOTROPY AND FLOW ON EQUILIBRIUM}

To completely determine the equilibrium 
we choose the plasma density, the Mach function and the anisotropy function profiles to be peaked on the magnetic axis and vanishing on the plasma boundary.  Specifically, for the Solovev solution we choose: $\widetilde{\rho} (\widetilde{u})=\widetilde{\rho} _a\left(1-\frac{\widetilde{u}}{\widetilde{u}_b}\right)^{1/2}$, $M_p^2(\widetilde{u})=M_{p_a}^2\left(1-\frac{\widetilde{u}}{\widetilde{u}_b}\right)^\mu$ and $\sigma _d(\widetilde{u})=\sigma _{d_a}\left(1-\frac{\widetilde{u}}{\widetilde{u}_b}\right)^n$
with $\widetilde{\rho}_a$ and $\widetilde{u}_b$ constant quantities, while for the Hernegger-Maschke solution we choose: $\widetilde{\rho}(\widetilde{u})=\widetilde{\rho}_a\left(\frac{\widetilde{u}}{\widetilde{u}_a}\right)^{1/2}$, $M_p^2(\widetilde{u})=M_{p_a}^2\left(\frac{\widetilde{u}}{\widetilde{u}_a}\right)^{\mu}$, $\sigma_d(\widetilde{u})=\sigma_{d_a}\left(\frac{\widetilde{u}}{\widetilde{u}_a}\right)^n$, with $\widetilde{\rho}_a$ and $\widetilde{u}_a$ constant quantities, respectively. It is noted  here that the above chosen  density function, peaked on the magnetic axis and vanishing on the boundary is  typical for tokamaks. Also, the Mach function adopted having a similar shape is reasonable at least in connection with experiments with  on axis focused  external momentum sources.
The functions $\widetilde{\rho}$, $M_p^2$ and $\sigma_{d}$ chosen depend on two free parameters; their maximum on axis and an exponent associated with the shape of the profile; the exponent of the function  $M_p^2$, connected with flow shear, is held fixed at $\mu =2$. 
 \newline
 The value of $M_{p_a}$ depends on the kind of tokamak  (conventional or spherical). On account of experimental evidence \cite{Brau,Pantis}, the toroidal rotation velocity in tokamaks is approximately $10^4-10^6\,ms^{-1}$ which for large conventional ones implies  $M_{p_a}^2 \sim 10^{-4}$, while the flow is stronger  for spherical tokamaks ($M_{p_a}^2 \sim 10^{-2}$) \cite{Menard}. In addition, from the requirement of positiveness for all pressures within the whole plasma region, we find that the pressure anisotropy parameter $\sigma_{d_a}$ takes higher values on spherical tokamaks than in the conventional ones, as shown on Table I;  also it must be $n\geq 2$.
\begin{table}
\begin{tabular}{|l||l|l||l|l|}
	\hline
	&\multicolumn{2}{l|}{Diamagnetic}&\multicolumn{2}{l|}{Paramagnetic}\\
	\cline{2-5}
	&ITER&NSTX&ITER&NSTX\\
	\hline\hline
	Parallel flow $(\lambda=0)$&0.08&0.11&0.089&0.12\\
	Non-parallel flow $(\lambda=0.5)$&0.10&0.13&0.094&0.13\\
	\hline
\end{tabular}
\caption{\emph{Approximate maximum permissible values of the free parameter $\sigma_{d_a}$ for the extended Solovev solution in connection with the non negativeness  of  pressure. }}
 \end{table}
  An argument why the flow and pressure anisotropy are stronger in spherical tokamaks  is that in this case the magnetic field is strongly inhomogeneous, as the  aspect ratio is too small. In contrast, for  the generalised Hernegger-Maschke equilibrium the pressure anisotropy takes a little higher values on ITER rather than on the NSTX-U tokamak, and this may be attributed to a peculiarity of this solution. \par
When the plasma is diamagnetic the toroidal magnetic field inside the plasma decreases from its vacuum value, and consequently the profile of the function $\widetilde{I}$ is expected to be hollow ($\widetilde{B}_\phi =\frac{\widetilde{I}}{\xi}$). As shown in Fig. (6), as $\sigma_{d_a}$ becomes larger the field increases, and for sufficient high $\sigma_{d_a}$ it becomes peaked on the magnetic axis. This means that increasing pressure anisotropy acts paramagnetically in terms of its maximum value on axis, $\sigma_{d_a}$. 
\begin{figure}
  \centering
    \includegraphics[width=3.5in]{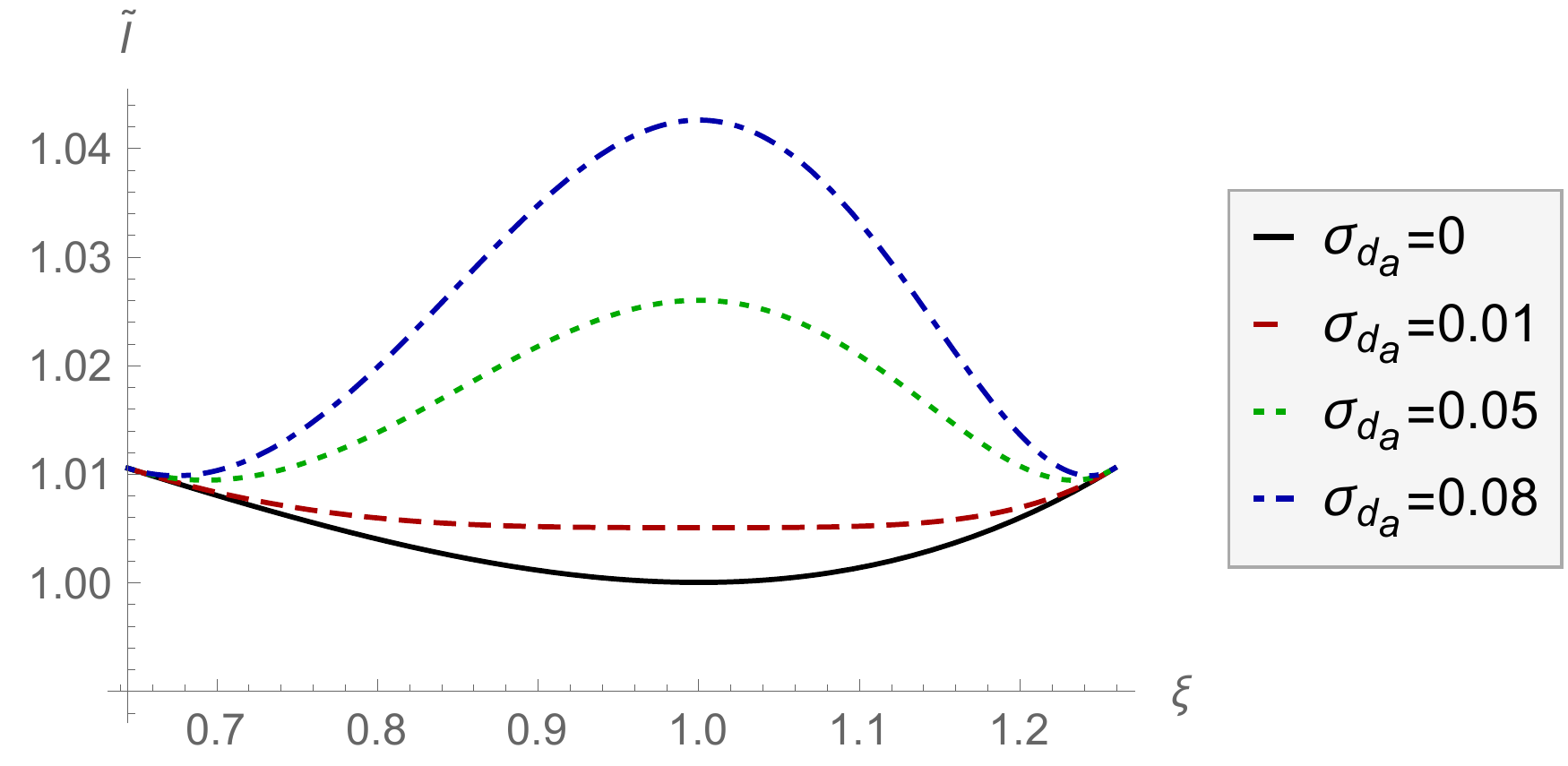}
     \caption{\emph{The paramagnetic action of pressure anisotropy through the parameter $\sigma_{d_a}$ on diamagnetic ITER-like equilibria with field-aligned flow, on the midplane $\zeta=0$, for the  extended Solovev solution. This result also holds for Hernegger-Maschke-like equilibria and paramagnetic plasmas, as well as for non-parallel flow.}}    
\end{figure}
Additionally, plasma flow through $M_{p_a}^2$ also acts paramagnetically, but its effects are weaker than that of pressure anisotropy, as shown in Fig. (7).
\begin{figure}
  \centering
    \includegraphics[width=3.5in]{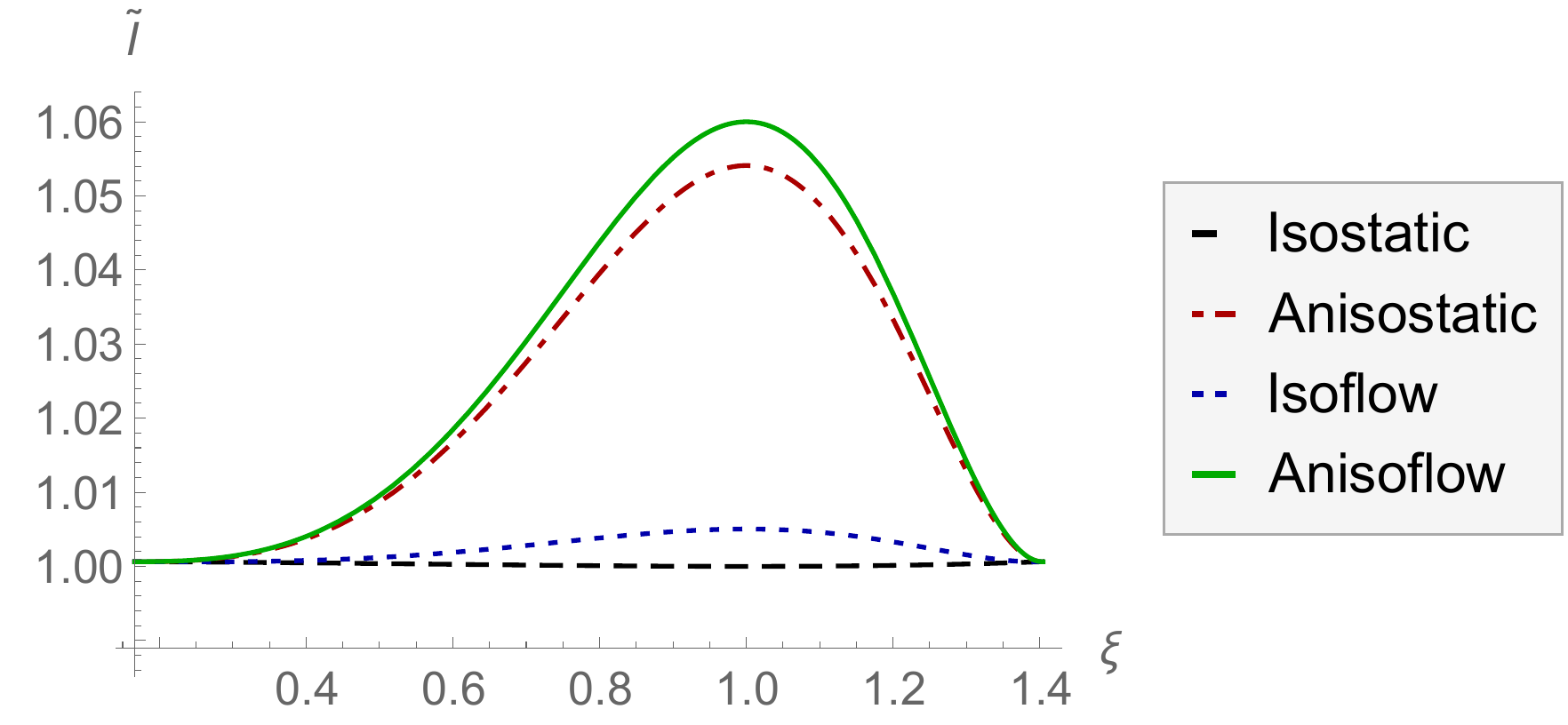}
     \caption{\emph{The additive paramagnetic action of anisotropy and flow on NSTX-like diamagnetic equilibria, on the midplane $\zeta=0$. We note that anisotropy (red-dashed-dotted curve) has a stronger impact than the flow (Blue Dotted curve) on equilibrium. The maximum paramagnetic action is found when both anisotropy and flow are present (green-straight curve). }} 
\end{figure}
On the other side, pressure anisotropy may also act diamagnetically through the shaping parameter $n$ when $\sigma_{d_a}$ is fixed [cf. Fig. (8)].
\begin{figure}
  \centering
    \includegraphics[width=3.5in]{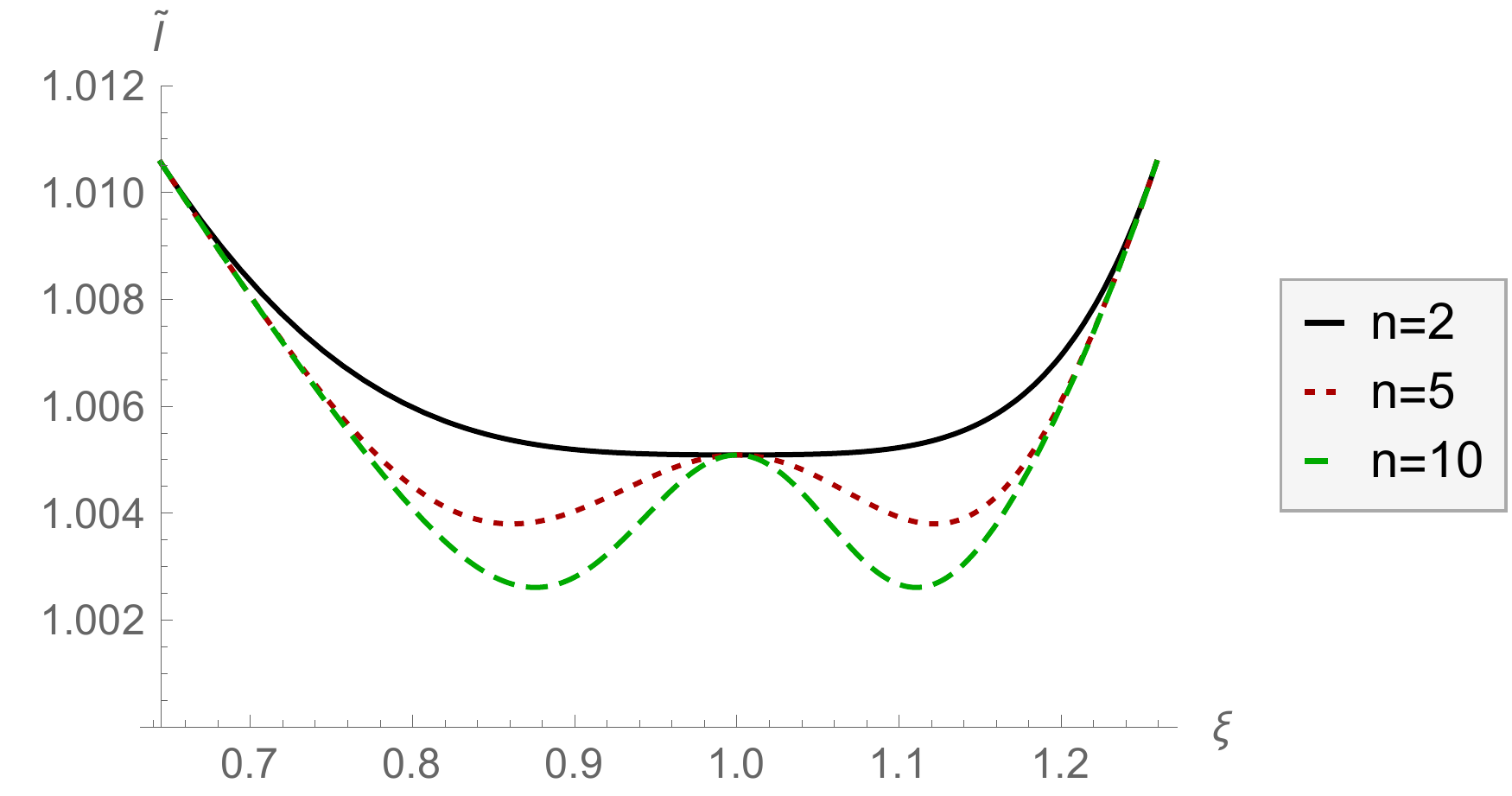}
     \caption{\emph{Raising the free parameter $n$ of the anisotropy function, decreases the toroidal magnetic field in the off-axis region, leading to a diamagnetic action.} }    
\end{figure} 
\par
For the extended diamagnetic Solovev solution, in the static and isotropic case  the toroidal current density monotonically increases from $\xi_{in}$ to $\xi_{out}$:
\beq \widetilde{J}_\phi =1.548\xi -\frac{0.333}{\xi}\eeq
When anisotropy is present, there are three regions where $\widetilde{J}_\phi$ displays different behavior: for $\xi_{in}<\xi<\xi_1$ and $\xi_2<\xi<\xi_{out}$ it decreases, while for $\xi_1<\xi<\xi_2$ it increases, compared with the isotropic case, as shown in Fig. (9). 
\begin{figure}
  \centering
    \includegraphics[width=3.5in]{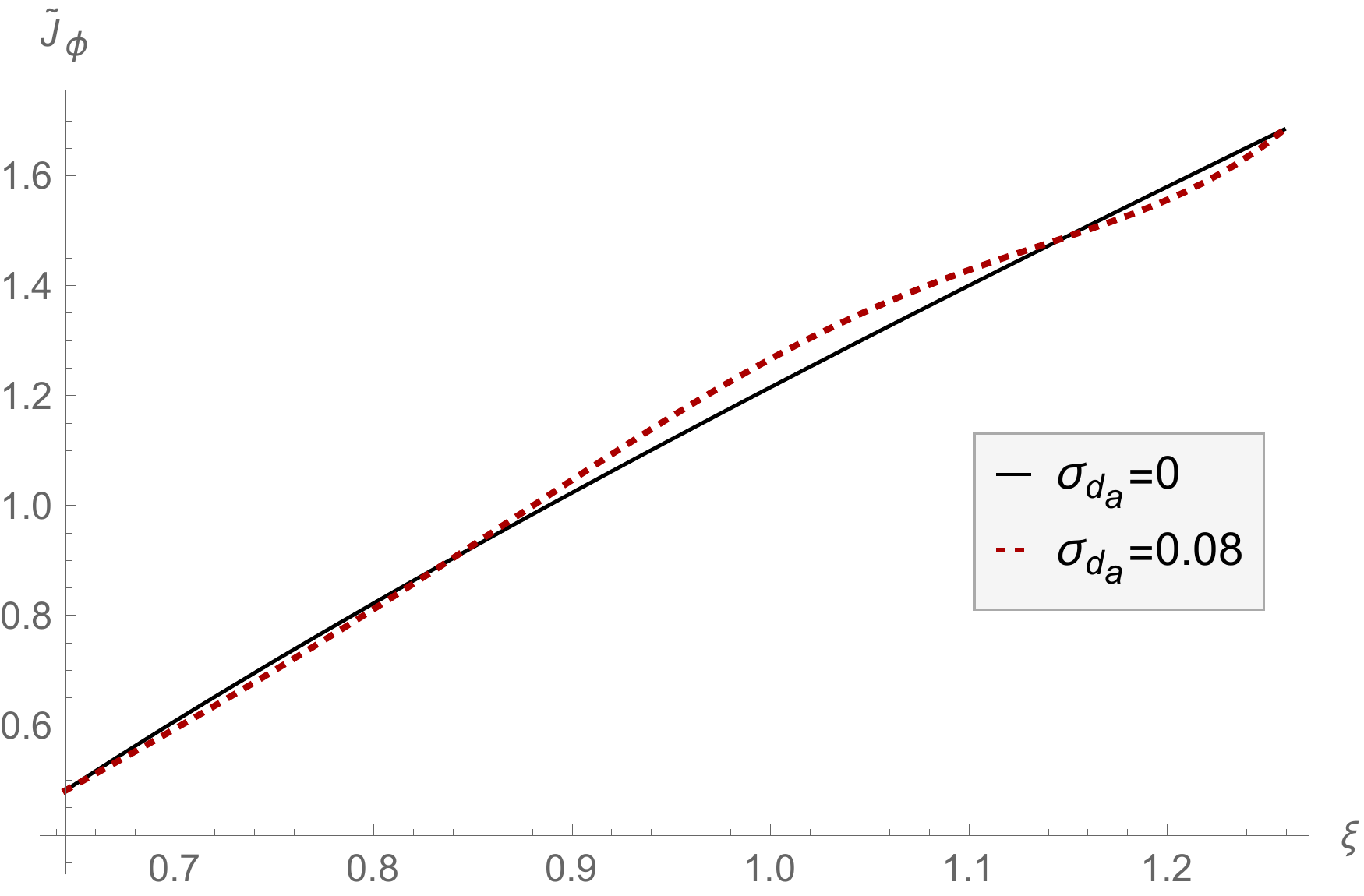}
     \caption{\emph{Diamagnetic ITER-like $\widetilde{J}_\phi(\sigma_{d_a})$ on the midplane $\zeta=0$, for $\lambda=0$, on the basis of the Solovev-like solution. For non-parallel flow the intersection points are displaced a little closer to the magnetic axis.}}  
      \end{figure}
      When the plasma is paramagnetic, $\widetilde{J}_\phi$ sharply falls off near the axis of symmetry, and then behaves diamagnetic-like. In contrast, the extended Hernegger-Maschke solution has a more realistic $\widetilde{J}_\phi$ profile peaked on the magnetic axis  and vanishing on the boundary.
In this case $\widetilde{J}_\phi$  slightly increases with anisotropy. 
\par
Furthermore, in the presence of pressure anisotropy the poloidal component $\widetilde{J}_\zeta$ of the Solovev and the radial $\widetilde{J}_\xi$ of the Hernegger-Maschke solution, present two extrema on the plane ($\zeta=\zeta_a$) containing the magnetic axis with their absolute values to be increasing with $\sigma_{d_a}$ [Fig. (10)]. For fixed $\sigma_{d_a}$, the higher $n$ is the closer to the magnetic axis  are located the extrema.
\begin{figure}
  \centering
    \includegraphics[width=3.75in]{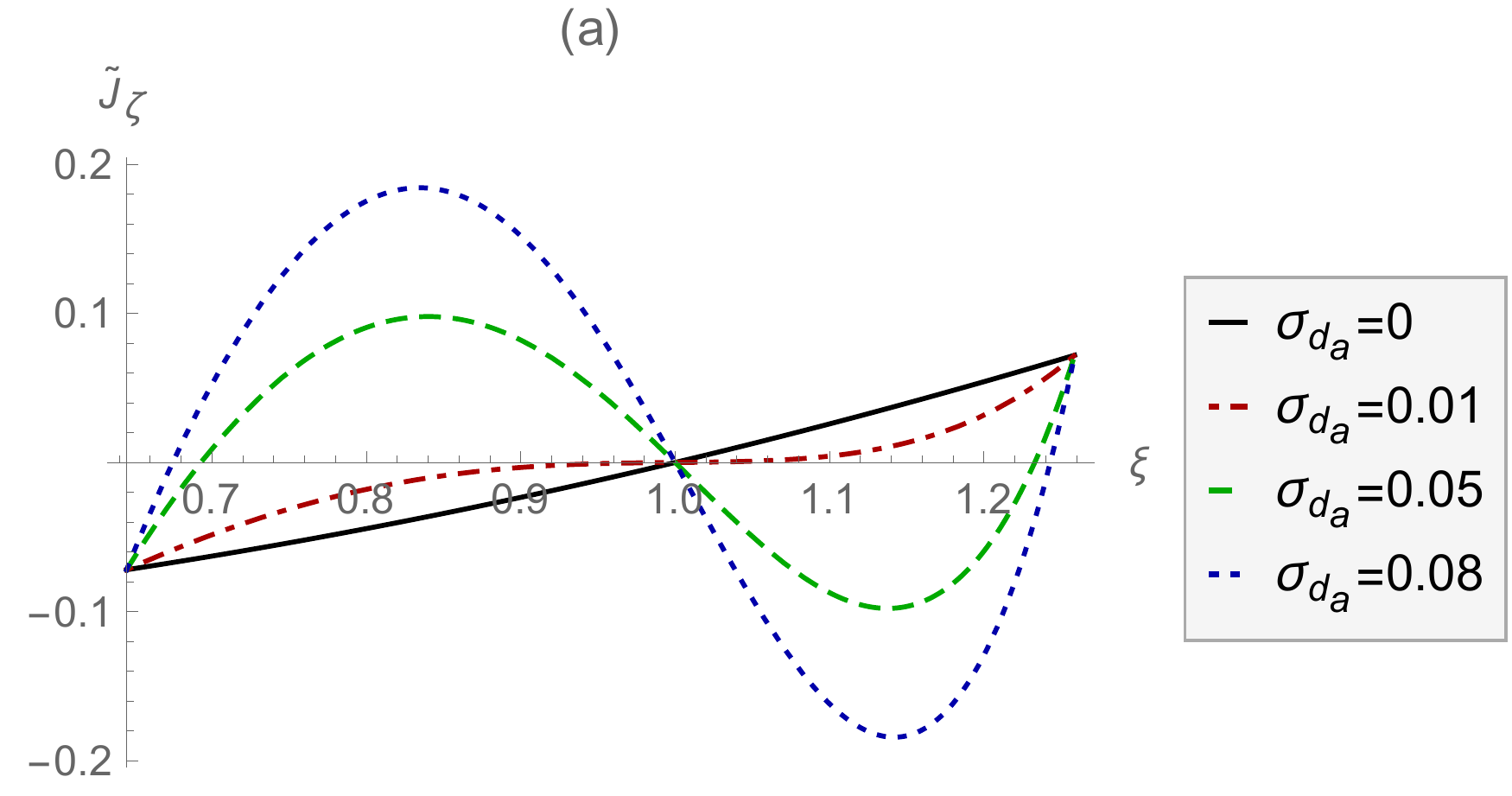}
    \includegraphics[width=3.75in]{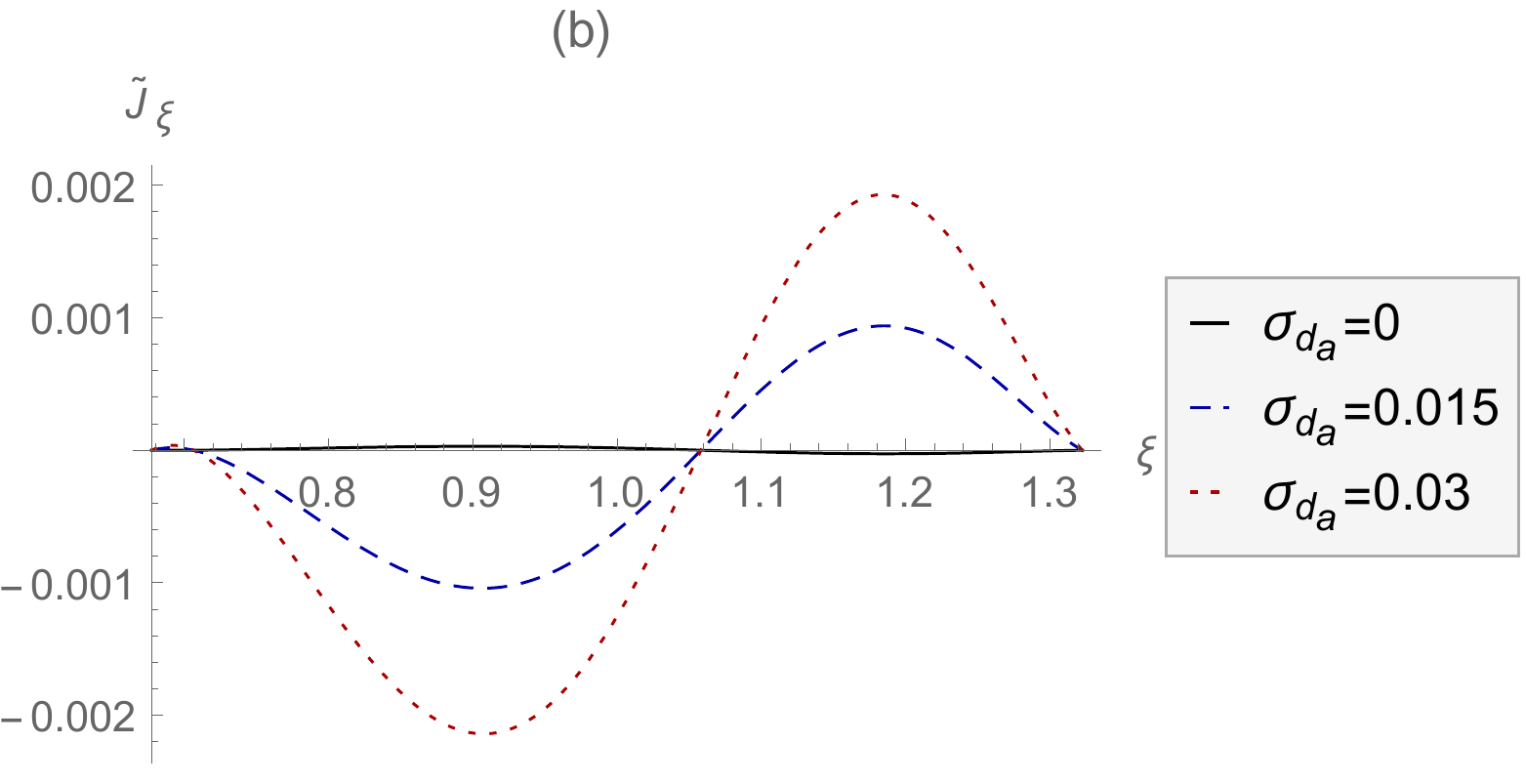}
     \caption{\emph{Variation of the poloidal components of current density for ITER-like diamagnetic equilibria in the presence of pressure anisotropy: (a) $\widetilde{J}_\zeta$ on the midplane $\zeta=0$ on the basis of the extended Solovev solution, (b) $\widetilde{J}_\xi$ on the plane $\zeta=\zeta_a$ on the basis of the extended Hernegger-Maschke solution. }}    
\end{figure} 
\newline
\newline
The  Solovev  toroidal velocity is expressed as
\beq \label{velocity Solovev}\widetilde{v}_\phi =\frac{\widetilde{I}}{\xi}\frac{M_p}{\sqrt{\widetilde{\rho}}}-\xi \sqrt{1-\frac{M_p^2}{1-\sigma _d}}\left(\frac{2\lambda \widetilde{\overline{p}}_{s_a}}{\widetilde{\rho}(1+\delta ^2)}\left(1-\frac{\widetilde{u}}{\widetilde{u}_b}\right)\right)^{1/2}\eeq
Thus, for parallel flow the second term in Eq. (\ref{velocity Solovev}) vanishes and $\widetilde{v}_\phi$ behaves like $\widetilde{I}$ as concerns its dependence on  $M_{p_a}^2$, $\sigma_{d_a}$ and $n$. We can see the increase of the maximum value of the toroidal velocity with $\sigma_{d_a}$, displaced on the left side of the magnetic axis, in Fig. (11), for an ITER-like diamagnetic configuration. For the NSTX the impact of anisotropy on $\widetilde{v}_{\phi}$ is qualitatively similar but quantitatively slightly stronger because of the higher values of $\sigma_{d_a}$. This behavior holds for the Hernegger-Maschke-like solution. For non parallel flow   $\widetilde{v}_\phi$ changes sign because of the negative second term in Eq. (\ref{velocity Solovev}).
\begin{figure}
  \centering
    \includegraphics[width=3.5in]{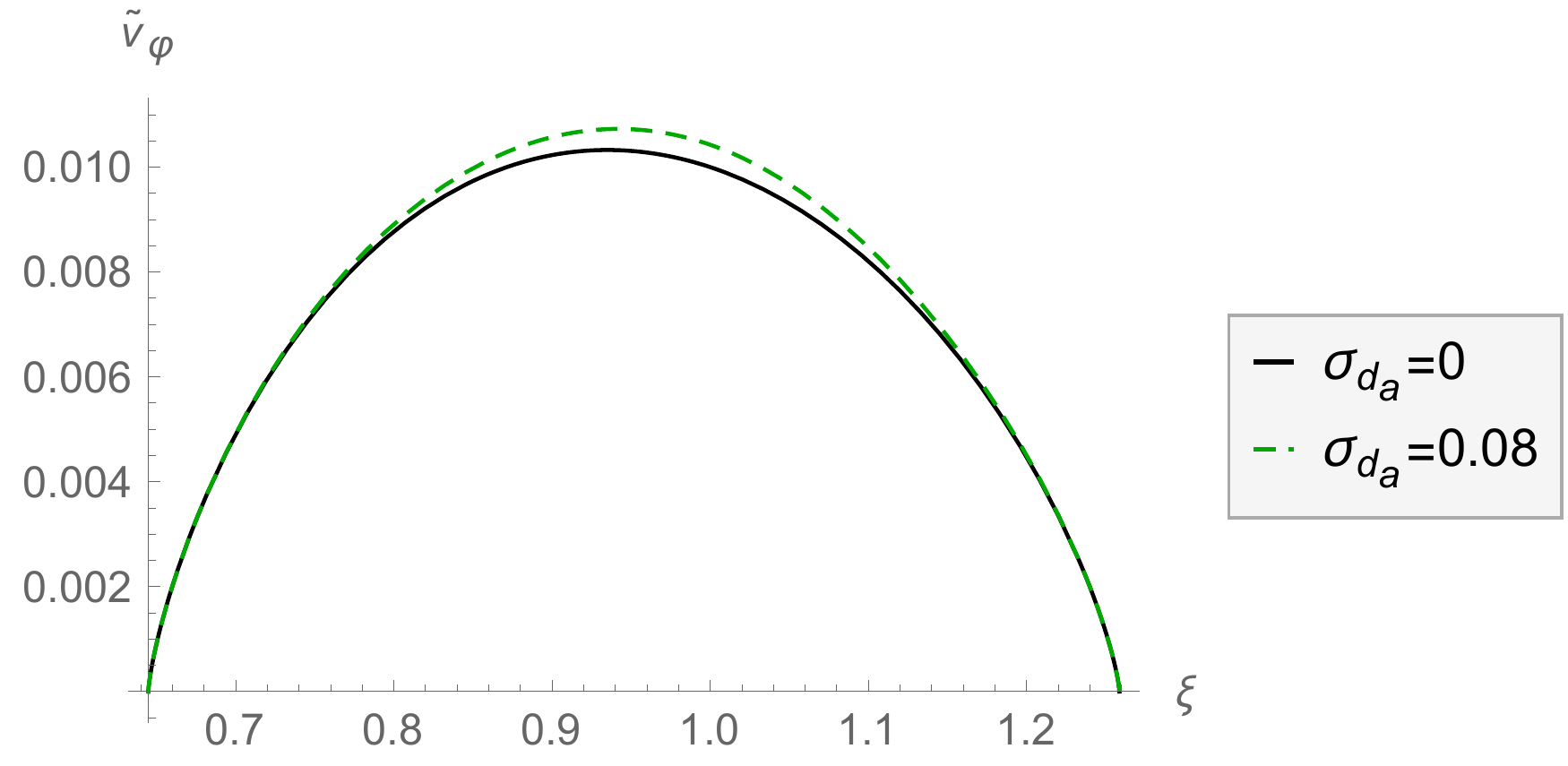}
     \caption{\emph{Diamagnetic ITER-like  $\widetilde{v}_\phi(\sigma_{d_a})$ profile for $\lambda=0$.}}    
\end{figure} 
\par
When the plasma is paramagnetic $\widetilde{v}_\phi$ reverses near the axis of symmetry and then behaves as the diamagnetic one to the right of the reversal point, as shown in Fig. (12). In spherical tokamaks the reversal point is displaced closer to the magnetic axis and $\widetilde{v}_\phi$ remains positive in a larger region than in the conventional ITER-like one. Reversal of $\widetilde{v}_\phi$ during the transition to improved confinement regimes have been observed in ASDEX Upgrade \cite{McD} and in LHD \cite{Ida}.
\begin{figure}
  \centering
    \includegraphics[width=3.5in]{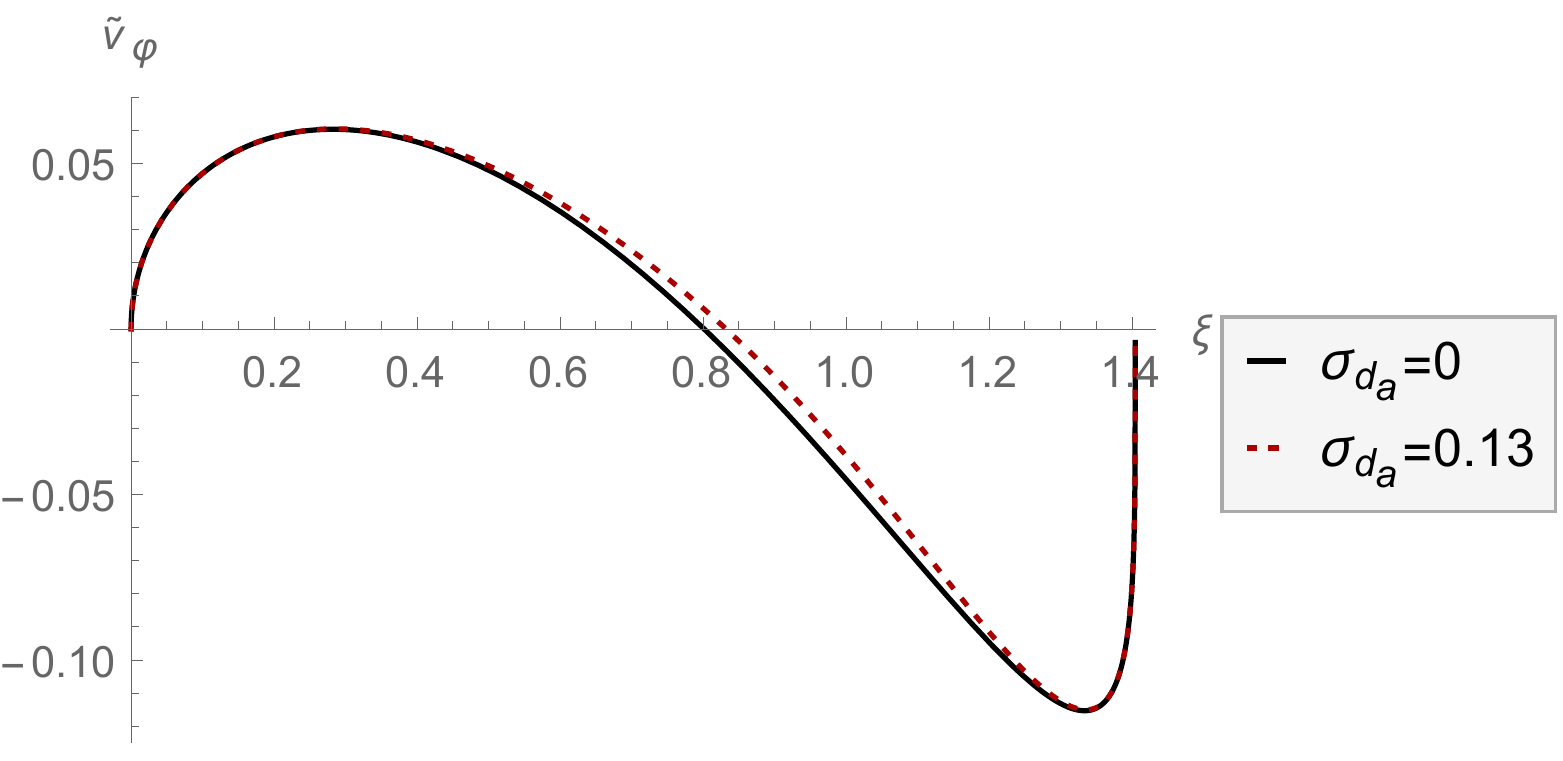}
     \caption{\emph{Paramagnetic $\widetilde{v}_\phi$ on the midplane $\zeta=0$ for NSTX-like equilibria with non-parallel flow for $\lambda=0.5$.} }
     \end{figure}
    
     \par{
Pressure anisotropy has an appreciable impact on the various pressures, with $\widetilde{\overline{p}}_{\parallel}$ increasing, while $\widetilde{\overline{p}}_{\bot}$ and $<\widetilde{p}>$ decreasing  with $\sigma_d$ as expected by Eqs. (\ref{average pressure})-(\ref{parallel pressure}). For a Solovev-like diamagnetic equilibrium the ratio of the scalar pressures parallel and perpendicular to the magnetic field is approximately equal for the two kinds of tokamak: {\tiny$\left(\frac{\widetilde{p}_{\parallel}}{\widetilde{p}_{\bot}}\right)_{ITER}\approx 1.227 $\normalsize},  {\tiny$\left(\frac{\widetilde{p}_{\parallel}}{\widetilde{p}_{\bot}}\right)_{NSTX}\approx 1.099$\normalsize}. In addition, the ratio of the maximum values of the average pressures for these two tokamaks is {\tiny$\frac{<\widetilde{p}>_{NSTX}}{<\widetilde{p}>_{ITER}}\approx 2.17$\normalsize}. For a Hernegger-Maschke-like diamagnetic equilibrium, the respective ratios are: 
\tiny$\left(\frac{\widetilde{p}_{\parallel}}{\widetilde{p}_{\bot}}\right)_{NSTX-U}\approx 1.08$ \normalsize  
 and \tiny$\left(\frac{\widetilde{p}_{\parallel}}{\widetilde{p}_{\bot}}\right)_{ITER}\approx 1.5$\normalsize. Also, for this equilibrium we found  \tiny$\frac{<\widetilde{p}>_{NSTX-U}}{<\widetilde{p}>_{ITER}}\approx 2.73$\normalsize, a ratio that approaches the respective Solovev one.}
\par
As  expected by Eqs. (\ref{perpendicular pressure}) and (\ref{parallel pressure}) the flow has a slightly stronger impact on $\widetilde{\overline{p}}$ than pressure anisotropy, as also shown in Fig. (13). At last, the rest of the equilibrium quantities and confinement figures of merit as the local toroidal  beta on axis and the safety factor are almost insensitive to anisotropy.

\begin{figure}
  \centering
    \includegraphics[width=3.5in]{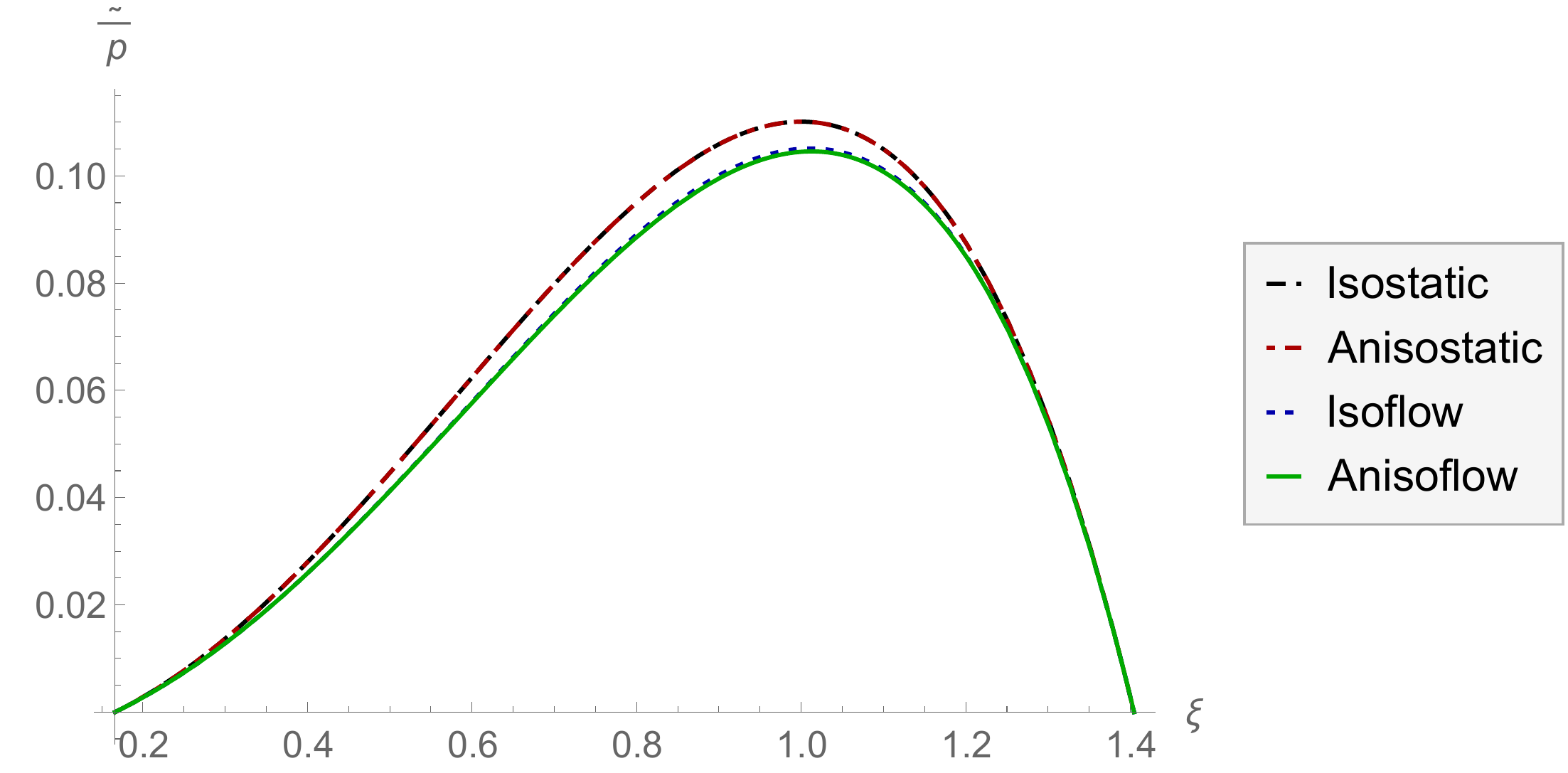}
     \caption{\emph{The influence of pressure anisotropy against the flow on $\widetilde{\overline{p}}$ on the midplane $\zeta=0$, for the NSTX diamagnetic equilibria with parallel flow ($\lambda=0$). When the flow is present the overall effective pressure decreases from its static value, while the presence of pressure anisotropy does not have an important effect on it. The maximum attainable values for the parameters $M_{p_a}^2$ and $\sigma_{d_a}$  used are imposed by the non negativeness of pressure.} }
\end{figure}

\section{CONCLUSIONS}
A generalised Grad-Shafranov equation [Eq. (\ref{GGS psi})] governing axisymmetric plasma equilibria in the presence of pressure anisotropy and incompressible flow was derived. This equation recovers known GS-like equations governing static anisotropic equilibria and isotropic equilibria with plasma flow. Also for static isotropic equilibria the equation is reduced to the usual well known GS equation. The derivation was based 
on 
 a diagonal pressure tensor with one element parallel to the magnetic field, $p_{\parallel}$, and two equal perpendicular ones, $p_{\bot}$. As a measure of the pressure anisotropy we introduced the function $\sigma_d=\mu_0\frac{p_{\parallel}-p_{\perp}}{B^2}$, assumed to be uniform on magnetic surfaces, while the flow was expressed by the poloidal Alfv\' enic Mach function $M_{p}=\frac{v_{pol}}{v_{A_{pol}}}$, where $v_{A_{pol}}$ is the Alfv\' en velocity. The form of the equation containing the sum $M_{p}^2+\sigma_d$ indicates that pressure anisotropy and flow act additively with the only exception the electric field term. In addition we derived a generalised Bernoulli equation [Eq. (\ref{Bernoulli})] involving the effective isotropic pressure $\overline{p}=\frac{p_{\parallel}+p_{\bot}}{2}$. \par
On the basis of a simpler form of the GGS equation obtained by a generalised transformation,  the transformed equation was linearised and solved for appropriate choices of the free functions appearing in it. Specifically, an extended Solovev solution describing configurations with the plasma boundary coinciding with a seperatrix, and an extended Hernegger-Maschke solution with a fixed boundary possessing an X-point imposed by appropriate boundary conditions, were derived. Employing these  solutions, ITER, NSTX and NSTX-U-like equilibria for arbitrary flow, both diamagnetic and paramagnetic, were constructed. In addition, we examined  the  impact of anisotropy -through the parameters $\sigma_{d_a}$ and $n$, defining the maximum value and the shape of the function $\sigma_d$- and flow -through the Alfv\' enic Mach number $M_{p_a}^2$ defining the maximum of the function $M_p^2$- on the equilibria constructed and came to the following conclusions. \par
Pressure anisotropy has a stronger impact on equilibrium than that of the flow because the maximum permissible values of $\sigma_{d_a}$ are in general higher than the respective $M_{p_a}^2$ ones, with the effects of the flow to be more noticeable in the spherical tokamaks. In addition, both anisotropy and flow through the parameters $\sigma_{d_a}$ and $M_{p_a}^2$ have an additive paramagnetic impact on equilibrium, with stronger  paramagnetic effects  in  spherical tokamaks, while anisotropy through $n$ acts diamagnetically. Furthermore, pressure anisotropy has an appreciable impact on equilibrium quantities such as the current density, the toroidal velocity and the parallel and perpendicular pressures, while $\overline{p}$ is slightly affected by the pressure anisotropy and more by the  flow.\par

On the   basis of the GGS obtained in this study one can develop a code to solve the problem for arbitrary   choices of the free functions involved  in order to  deal with  experimental  equilibrium profiles  or  extend existing codes, e.g. the isotropic HELENA code for incompressible parallel flows \cite{HELENA}. Also, it is interesting to extend the papers on static equilibria with reversed current density \cite{Wang}-\cite{Martins} in the presence of incompressible flow and pressure anisotropy.  In addition, the study can be extended for the more general case of helically symmetric equilibria. 

Let us finally note that complete understanding of the equilibrium with plasma flow and pressure anisotropy requires substantial additional work in connection with compressibility, alternative potentially more  pertinent  physical assumptions on the functional dependence of the anisotropy function $\sigma_d$ and more realistic numerical solutions. However, in these cases   the  reduced equilibrium equations are expected to  be much more complicated compared with the relative simple GGS derived in the present study which contributes to understanding  the underlying physics.

\section*{ACKNOWLEDGEMENTS}
One of the authors (GNT) would like to thank  Henri Tasso, 
George Poulipoulis and Apostolos Kuiroukidis for very useful discussions.
This work has been carried out within the framework of the EUROfusion Consortium and has received funding from  (a) the National Programme for the Controlled Thermonuclear Fusion, Hellenic Republic, (b) Euratom research and training programme 2014-2018 under grant agreement No 633053. The views and opinions expressed herein do not necessarily reflect those of the European Commission.
\newpage
\appendix

 \section{\\Details for the construction of the diverted equilibrium of Fig. (4)} \label{App:APPENDIX A}
 
 In order to make the analysis more convenient, we factorize (\ref{HM u}) with respect to the coefficient $a_1$, so that the system of algebraic equations that will be derived  from the imposed boundary conditions to be inhomogeneous and therefore easier to be solved numerically:
 
 {\small
 	\begin{align}
 	\widetilde{u}& =a_1\left\{ M_{\nu _1,\frac{1}{2}}(\varrho)cos(\zeta)+\frac{b_1}{a_1}M_{\nu _1,\frac{1}{2}}(\varrho)sin(\zeta)+\frac{c_1}{a_1}W_{\nu _1,\frac{1}{2}}(\varrho)cos(\zeta)+\frac{d_1}{a_1}W_{\nu _1,\frac{1}{2}}(\varrho)sin(\zeta)\right. \nonumber \\
 	&\left. {}+\sum _{j=2}^N \left[\frac{a_j}{a_1} M_{\nu _j,\frac{1}{2}}(\varrho)cos(j\zeta)+\frac{b_j}{a_1}M_{\nu _j,\frac{1}{2}}(\varrho)sin(j\zeta)+\frac{c_j}{a_1}W_{\nu _j,\frac{1}{2}}(\varrho)cos(j\zeta)+\frac{d_j}{a_1}W_{\nu _j,\frac{1}{2}}(\varrho)sin(j\zeta)\right]\right\}
 	\end{align}
 	\normalsize}
 
 Now, by setting $a_j^*=\frac{a_j}{a_1}$, $b_j^*=\frac{b_j}{a_1}$, $c_j^*=\frac{c_j}{a_1}$, and $d_j^*=\frac{d_j}{a_1}$, then the solution can be expressed as
 \beq \widetilde{u}(\varrho ,\zeta)=a_1\widetilde{u}^*(\varrho ,\zeta)\eeq
 where
 {\small
 	\beq \label{HM u star}\widetilde{u}^*(\varrho ,\zeta)=\sum _{j=1}^N \left[a_j^*M_{\nu _j,\frac{1}{2}}(\varrho)cos(j\zeta)+b_j^*M_{\nu _j,\frac{1}{2}}(\varrho)sin(j\zeta)+c_j^*W_{\nu _j,\frac{1}{2}}(\varrho)cos(j\zeta)+d_j^*W_{\nu _j,\frac{1}{2}}(\varrho)sin(j\zeta)\right] \eeq
 	\normalsize}
 with $a_1^*=1$. \par
 An up-down asymmetric boundary consisting of a smooth upper part and a lower part that possesses an X-point, is described by the parametric equations introduced in \cite{Parametric}, with its boundary  being represented  by the following characteristic  points shown in Fig. (14):
 \begin{center}
 	Inner point : $(\xi_{in}=1-\frac{\alpha}{R_0},\zeta_{in}=0)$
 \end{center}
 \begin{center}
 	Outer point : $(\xi_{out}=1+\frac{\alpha}{R_0},\zeta_{out}=0)$
 \end{center}
 \begin{center}
 	Upper point : $(\xi_{up}=1-t\frac{\alpha}{R_0},\zeta_{up}=\kappa\frac{\alpha}{R_0})$
 \end{center}
 \begin{center}
 	Lower X-point : $(\xi_{x}=1+\frac{\alpha}{R_0}cos[\pi -tan^{-1}(\frac{\kappa}{t})],\zeta_{x}=-\kappa\frac{\alpha}{R_0})$
 \end{center}
 \begin{figure}
 	\centering
 	\includegraphics[width=3.5in]{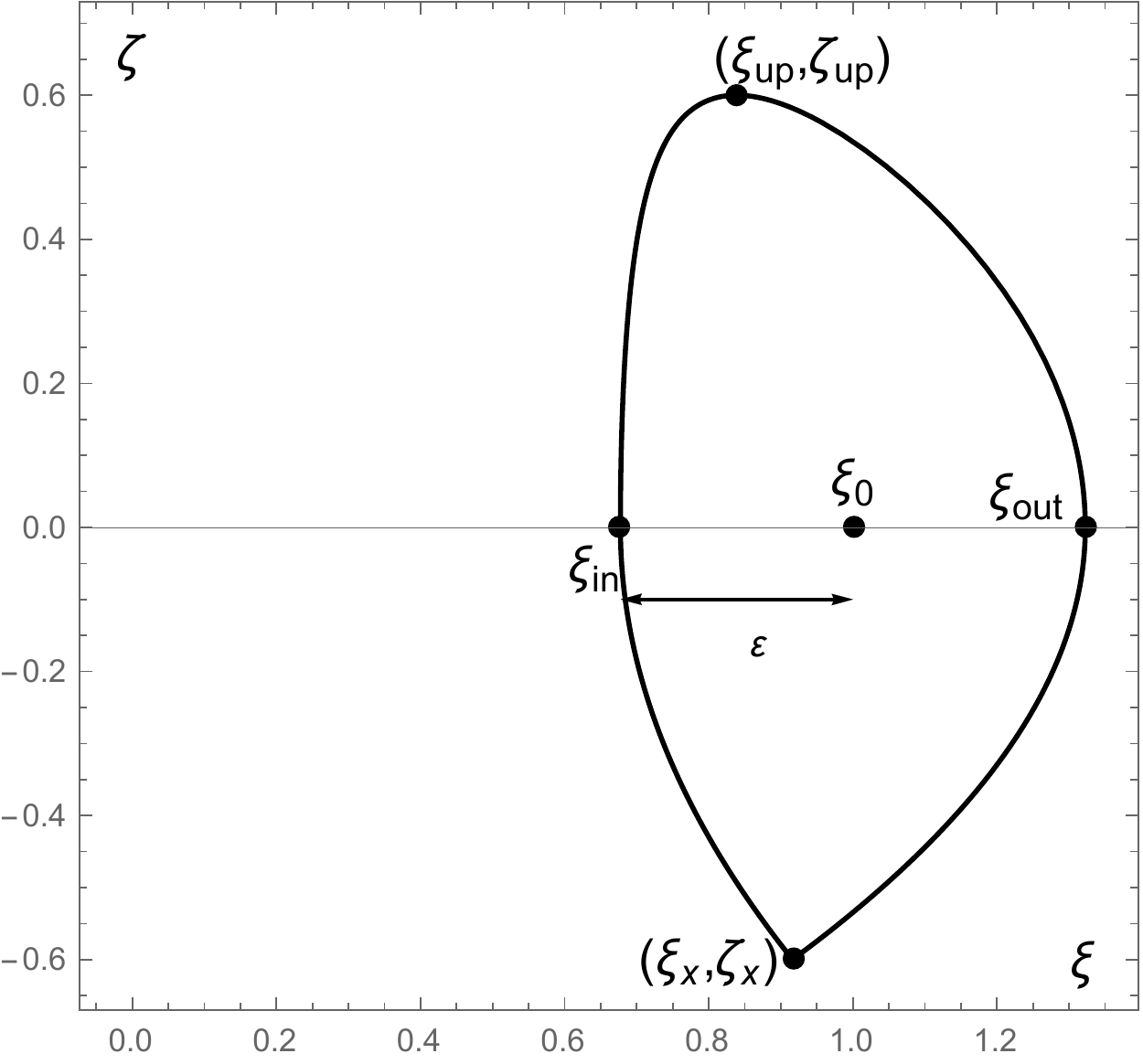}
 	\caption{\textit{Characteristic points determining an up-down asymmetric boundary, described by the parametric equations given in \cite{Parametric}.}} 
 \end{figure}
 In order to calculate the unknown coefficients of the solution we will impose the condition that $\widetilde{u}^*$ vanishes on the boundary. Function $\widetilde{u}^*$ is in general complex, and since it satisfies the GGS equation, then both its real and imaginary parts are also solutions of this equation. Here, following Ref. \cite{Guazzoto} we will work with the imaginary part of the flux function. So the first four conditions are:
 \beq Im[\widetilde{u}^*(\xi _{in},\zeta _{in})]=Im[\widetilde{u}^*(\xi _{out},\zeta _{out})]=Im[\widetilde{u}^*(\xi _{up},\zeta _{up})]=Im[\widetilde{u}^*(\xi _x,\zeta _x)]=0\eeq
 
 In addition, five  boundary conditions related with the first derivative of these  characteristic points are imposed: 
 \beq Im[\widetilde{u}_\zeta ^*(\xi _{in},\zeta _{in})]=Im[\widetilde{u}_\zeta ^*(\xi _{out},\zeta _{out})]=Im[\widetilde{u}_\xi ^*(\xi _{up},\zeta _{up})]=Im[\widetilde{u}_\zeta ^*(\xi _x,\zeta _x)]=Im[\widetilde{u}_\xi^*(\xi _x,\zeta _x)]=0\eeq
 where, $\widetilde{u}_\zeta^*=\frac{\partial \widetilde{u}^*}{\partial \zeta}$, and $\widetilde{u}_\xi^*=\frac{\partial \widetilde{u}^*}{\partial \xi}$. The above conditions guarantee  smoothness of the curve at the characteristic points; particularly, the curve is imposed to  be perpendicular to the midplane. Furthermore, there exist three other conditions  introduced in \cite{Cerfon}, that involve the second derivatives of $\widetilde{u}^*$ related with the curvature of the boundary curve in the characteristic points. These are:
 \beq Im[\widetilde{u}_{\xi \xi}^*(\xi _{up},\zeta _{up})]=\frac{\kappa}{\varepsilon cos^2w_1}Im[\widetilde{u}_\zeta ^*(\xi _{up},\zeta _{up})]\eeq
 \beq  Im[\widetilde{u}_{\zeta \zeta}^*(\xi _{in},\zeta _{in})]=-\frac{(1-w_1)^2}{\varepsilon \kappa ^2}Im[\widetilde{u}_\xi ^*(\xi _{in},\zeta _{in})]\eeq
 \beq Im[\widetilde{u}_{\zeta \zeta}^*(\xi _{out},\zeta _{out})]=\frac{(1+w_1)^2}{\varepsilon \kappa ^2}Im[\widetilde{u}_\xi ^*(\xi _{out},\zeta _{out})]\eeq
 where the parameter $w_1$ relates to the triangularity of the boundary, $sinw_1=t$.
 Thus, for ITER-like characteristics, by setting $j_{max}=4$ and choosing the free parameters $\widetilde{p}_2=19.5$, $\widetilde{X}_1=-0.3$, by the imposition of the above conditions we find the values for the unknown coefficients presented on Table II.

 \begin{table}[ht]
 
 \centering 
 \begin{tabular}{c c c c}
 \hline\hline
 & Coefficient Values  \\ 
 \hline
 $a_1^*$ & 1 & $c_1^*$ & 2.2319  \\
$a_2^*$ & -0.625449 & $c_2^*$ & -1.82077  \\
$a_3^*$ & 0.13783 & $c_3^*$ & 0.52774  \\
$a_4^*$ & -0.0166821 & $c_4^*$ & -0.0657547  \\
$b_1^*$ & 11.2021 & $d_1^*$ & 29.2229  \\
$b_2^*$ & -4.93763 & $d_2^*$ & -20.6201  \\
$b_3^*$ & 1.23997 & $d_3^*$ & 10.462  \\
$b_4^*$ & -0.117724 & $d_4^*$ & -2.64276  \\  [1ex]
 \hline
 \end{tabular}
 \caption{\emph{Values of the coefficients of the solution $\widetilde{u}^*$ for an ITER-like diamagnetic configuration. }}
 \end{table}
 
 	Once the solution $\widetilde{u}^*(\rho ,\zeta)$ is fully determined, we can find the position of the magnetic axis, by solving the equations $Im[\widetilde{u}_\xi^*]=0$ and $Im[\widetilde{u}_\zeta^*]=0$  located outside of the midplane $\zeta=0$ at $(\xi_a=1.05815,\zeta_a=0.0159088)$. Subsequently, with the aid of Eq. (\ref{safety factor on axis}) we impose the condition $q_a=1.1$, just for the Kruskal-Shafranov limit to be satisfied, implementation of which gives $a_1=1.07751$. Thus, solution $\widetilde{u}$ is fully determined, with its value on axis to be $\widetilde{u}_a=-0.0416752$.  Closed magnetic surfaces associated with $\widetilde{u}$-contours of the equilibrium configuration are shown in Fig. (4).


\section*{REFERENCES}

\begin{thebibliography}{99}
\bibitem{zonal} P. H. Diamond, S. I. Itoh, K. Itoh and T. S. Hahm, Plasma Phys. Control. Fusion 47 (2005) R35-R161
\bibitem{flow} K. G. McClements and M. J. Hole, Phys. Plasmas 17, (2010) 082509
\bibitem{CGL} G. F. Chew, M. L. Goldberger, F. E. Low, Proc. R. Soc. London 236 (1956) 112
\bibitem{so} L. S. Solovev, Sov. Phys. JETP 26, (1968) 400 
\bibitem{hema} F. Hernegger, in: E. Canobbio et al. (Eds.), Proceedings of the 5th Conference on Control.
Fusion, Vol. I Commissariat a l' Energie Atomique, Grenoble, 1972, p. 26.; E.K. Maschke,
Plasma Phys. 15, (1973) 535 
\bibitem{moso} A. I. Morozov and L. S. Solov{\rq}ev, 
	Rev. Plasma Phys. 8 (1980) 1
\bibitem{ha} E. Hameiri, Phys. Fluids 26 (1983) 230
\bibitem{T-Th}H. Tasso, G. N. Throumoulopoulos, Phys. Plasmas 5 (1998) 2378
%
\bibitem{couhi}R. Courant, D. Hilbert, Methods of mathematical physics
(Interscience Publishers, 1966), Vol. 2, p. 372
%
\bibitem{Cotsaftis} C. Mercier, M. Cotsaftis, Nucl. Fusion 1 (1961) 121
\bibitem{Clement} R. A. Clemente, Nucl. Fusion 33 (1993) 963
\bibitem{zwi} W. Zwingmann, L.G. Eriksson and P. Stubberfield, Plasma Phys. Control. Fusion 43 (2001) 1441
\bibitem{pu} V. D. Pustovitov, Plasma Phys. Control. Fusion 52 (2010) 065001
\bibitem{lepu} N. D. Lepikhin and V. D. Pustovitov, Plasma Physics Reports,  39 (2013) 605
\bibitem{fu} M. Furukawa, Phys. Plasmas 21 (2014) 012511 
\bibitem{Kuznetsov} E. A. Kuznetsov, T. Passot, V. P. Ruban, P. L. Sulem, Phys. Plasmas 22 (2015) 042114
%
\bibitem{iabo} R. Iacono, A. Bondeson, F. Troyon, and R. Gruber, Phys. Fluids B 2 (1990) 1794 
\bibitem{il} V. I. Ilgisonis, Phys. Plasmas 3 (1996) 4577 
\bibitem{gube} L. Guazzotto, R. Betti, J. Manickam, and S. Kaye, Phys. Plasmas 11, (2004) 604 
\bibitem{clst} R. A. Clemente and D.  Sterzo Plasma Phys. Control. Fusion 51 (2009) 085011
\bibitem{pu12} V. D. Pustovitov, AIP Conf. Proc. 1478  (2012) 50
\bibitem{qufi} Z S Qu, M Fitzgerald and M J Hole, Plasma Phys. Control. Fusion 56 (2014) 075007
\bibitem{ivma} A. A. Ivanov, A. A. Martynov, S. Yu. Medvedev and Yu. Yu. Poshekhonov, Plasma Physics Reports, 41 (2015)  203
%
\bibitem{pothta} G. Poulipoulis, G. N. Throumoulopoulos, H. Tasso, Phys. Plasmas J12, (2005) 042112 
\bibitem{Palumbo} D. Palumbo, Nuovo Cimento B 53 (1968) 507
\bibitem{PaBo} D. Palumbo and M. Balzano, Atti Acc. Sc. Lett. Palermo IV(I) (1983-84) 129 
\bibitem{Simintzis} C. Simintzis, G. N. Throumoulopoulos, G. Pantis, H. Tasso, Phys. Plasmas 8 (2001) 2641
\bibitem{Fr} J. P. Freidberg, in Ideal Magnetohydrodynamics (Plenum
Press, 1987), ps. 108 and 117
\bibitem{Menard}J. E. Menard, M. G. Bell, R. E. Bell, et al., Nucl. Fusion 43 (2003) 330
\bibitem{ITERbeta} T. C. Luce, C. D. Challis, S. Ide, et al., Nucl. Fusion 54 (2014) 013015
\bibitem{Arapoglou} I. Arapoglou, G. N. Throumoulopoulos, H. Tasso, Phys. Letters A (2013) 310
\bibitem{Lao} L. L. Lao, S. P. Hirshman and R. M. Wieland, Phys. Fluids 24 (1981) 1431
\bibitem{Bizarro} Joao P. S. Bizarro, Nucl. Fusion 54 (2014) 083015
\bibitem{mizu} Kishore Mishra, H. Zushi, H. Idei, et al.,  Nucl. Fusion 55 (2015) 083009
\bibitem{Brau} K. Brau, M. Bitter, R. J. Goldston et al., Nucl. Fusion 23 (1983) 1643
\bibitem{Pantis} G. N. Throumoulopoulos, G. Pantis, Phys. Fluids B 1 (1989) 1827
\bibitem{McD} R.M. McDermott, C. Angioni, G. D. Conway, et al., Nucl. Fusion 54 (2014) 043009
\bibitem{Ida} K. Ida, H. Lee, K. Nagaoka, et al., PRL 111 (2013) 055001
\bibitem{HELENA} G. Poulipoulis, G.N. Throumoulopoulos, C. Konz and EFDA ITM-TF contributors, ``Extending HELENA to equilibria with incompressible parallel plasma rotation'', 39th EPS Conference and 16th Int. Congress on Plasma Physics, Stockholm (2012) P4.027
\bibitem{Wang} S. Wang, Phys. Rev. Lett. 93 (2004) 155007
\bibitem{RoBiz95} P. Rodrigues and Joao P. S. Bizarro, Phys. Rev. Lett. 95
(2005) 015001
\bibitem{RoBiz99} P. Rodrigues and Joao P. S. Bizarro, Phys. Rev. Lett. 99
(2007) 125001
\bibitem{Martynov} A. A. Martynov, S. Yu. Medvedev, and L. Villard, Phys.
Rev. Lett. 91 (2003) 085004
\bibitem{Martins} Caroline G. L. Martins, M. Roberto, I. L. Caldas, and F.
L. Braga, Phys. Plasmas 18 (2011) 082508
\bibitem{Parametric} A. Kuiroukidis, G. N. Throumoulopoulos, Plasma Phys. Control. Fusion 57 (2015) 078001
\bibitem{Guazzoto} L. Guazzotto, J. P. Freidberg, Phys. Plasmas 14 (2007) 112508
\bibitem{Cerfon} A. J. Cerfon, J. P. Freidberg, Phys. Plasmas 17 (2010) 032502
\end{thebibliography}
 \end{document}